\documentclass[showpacs,preprintnumbers,amsmath,amssymb]{revtex4}

\usepackage{graphicx}
\usepackage{dcolumn}
\usepackage{bm}
\usepackage{url}
\usepackage{epsfig}
\usepackage{multirow}
\usepackage{tensor}

\def\be{\begin{equation}}
\def\ee{\end{equation}}
\def\ba{\begin{eqnarray}}
\def\ea{\end{eqnarray}}

\newcommand{\fr}[2]{\frac{#1}{#2}}
\def\eff{\rm{eff}}
\def\m{\rm{m}}
\def\N{\rm{N}}
\def\Omo{\Omega_{\rm{m}0}}
\def\Oma{\Omega_{\rm{m}}(a)}
\def\Omz{\Omega_{\rm{m}}(z)}
\def\w{\omega}
\def\obs{\rm{obs}}

\def\res{\rm{res}}

\def\Geff{G_{\rm{eff}}}
\def\GN{G_{\rm{N}}}

\newcommand{\JCAP}{J.\ Cosmol.\ Astropart.\ Phys.}

\def\ga{\mathrel{\raise.3ex\hbox{$>$\kern-.75em\lower1ex\hbox{$\sim$}}}}
\def\la{\mathrel{\raise.3ex\hbox{$<$\kern-.75em\lower1ex\hbox{$\sim$}}}}

\begin{document}

\title{Evidence for departure from $\Lambda$CDM with LSS}


\author{Seokcheon Lee}
\email{skylee@phys.sinica.edu.tw}
\affiliation{Institute of Physics, Academia Sinica, Nankang,
Taiwan 11529, R.O.C.}


\begin{abstract}
We investigate the growth index parameter $\gamma$ and the time variation of the gravitational constant $\Geff$ by using the currently available growth function $f(z)$ data at different redshifts, with and without scaling to the fiducial $\Lambda$CDM model. We inquire the four different models of $\gamma$ including a constant $\gamma$. From a $\chi^2$ minimization, we constrain the parameter spaces of models and show that $\Lambda$CDM model is excluded by 1-$\sigma$ level from current $f(z)$ data. $\Geff$ is different from the Newton's gravitational constant $\GN$ in modified gravity theories and interestingly, the current data shows that $\Geff \neq \GN$ at $z \gtrsim 0.2 \sim 0.3$ with 3-$\sigma$ level. From these, we conclude that Einstein's General Relativity with $\Lambda$CDM is ruled out by 99 \% confidence level from large scale structure observations.

\end{abstract}

\pacs{95.36.+x, 98.65.-r, 98.80.-k. }

\maketitle

\section{Introduction}
\setcounter{equation}{0}
Since the discovery of the current accelerating expansion of the Universe, both a dark energy (DE) and a modification of gravity (MG) at cosmological scales are the most commonly proposed candidates for explaining cosmic acceleration. It has been proposed that complementary observations of both geometrical tests and the cosmological evolution of the large scale structure (LSS) formation might reveal the origin of the cosmic acceleration among those models.

It has been known that the flat $\Lambda$CDM cosmological model which is named to the so-called concordance model of Big Bang cosmology is consistent with the majority of current cosmological observations including both background evolution (e.g. supernovae type Ia, cosmic microwave background, baryon acoustic oscillation, etc) and LSS growth function (e.g. redshift space distortions, growth rate of clustering, weak lensing, etc). We re-examine the consistency of the $\Lambda$CDM model by using the current growth function data. We also investigate any time variation of the effective gravitational constant from the same data.

\section{Constraints on Growth function}
\setcounter{equation}{0}

The sub-horizon scales linear perturbation of the matter ($\delta_{\m} = \delta \rho_{\m} / \rho_{\m}$) for many DE and MG models is governed by \be \ddot{\delta}_{\m} + 2 H \dot{\delta}_{\m} = 4 \pi G_{\eff} \rho_{\m} \delta_{\m} \, , \label{deltamt} \ee where dot means the derivatives with respect to cosmic time t, $\rho_{\m}$ is the mean matter density, $H$ is the Hubble parameter given by the Friedmann equation ($H^2 = \fr{8 \pi G_{\N}}{3} \rho_{cr}$), and $G_{\eff}$ is the effective gravitational constant obtained from the Poisson equation. $G_{\eff}$ becomes Newton's gravitational constant $G_{\N}$ in Einstein's general relativity. We change the variable in Eq. (\ref{deltamt}) from $t$ to $\ln a$ to obtain \be \fr{d^2 \ln \delta_{\m}}{d \ln a^2} + \Bigl( \fr{d \ln \delta_{\m}}{d \ln a} \Bigr)^2 + \Biggl[ \fr{1}{2} - \fr{3}{2} \omega \Bigl( 1 - \Omega_{\m}(a) \Bigr) \Biggr] \fr{d \ln \delta_{\m}}{d \ln a} = \fr{3}{2} \fr{G_{\eff}}{G_{\N}} \Omega_{\m}(a) \, , \label{deltama} \ee where $\omega$ is the effective equation of state of dark energy and $\Oma = \rho_{\m}(a) / \rho_{\rm{cr}}$.

The growth function, $f$ is defined as the logarithmic derivative of the linear growth factor, $D$ which is the growing mode solution of $\delta_{\m}$
\be f(a) \equiv \fr{d \ln D}{d \ln a} \simeq \Omega_{\m}(a)^{\gamma} \, , \label{f} \ee where $\gamma$ is the so-called growth index parameter. For $\w = -1/3$ or $-1$ dark energy models, there exit the exact analytic solutions of $\gamma$ at present epoch \cite{09051522} \be \gamma_{0} = \ln \Biggl[-\fr{3}{2} - \fr{3}{2} \omega (1 - \Omo) - 3 \w \Omo^{\fr{3}{2}} \fr{\Gamma[1-\fr{5}{6\w}]/\Gamma[-\fr{5}{6\w}]}{F[\fr{3}{2},-\fr{5}{6\w},1-\fr{5}{6\w},1-\Omo^{-1}]} \Biggr] \Bigl/ \ln \Omo \, , \label{gamma0} \ee where $\Gamma$ is the gamma function, $F$ is the hypergeometric function, and $\Omo$ is the present value of matter energy density contrast. Thus, for $\Lambda$CDM $\gamma_{0} = \ln \Bigl[-\fr{3}{2} \Omo + 3 \Omo^{\fr{3}{2}} \Gamma[\fr{11}{6}]/(\Gamma[\fr{5}{6}] F[\fr{3}{2}, \fr{5}{6}, \fr{11}{6}, 1-\Omo^{-1}]) \Bigr] \Bigl/ \ln[\Omo]$ and varies from $0.558$ to $0.553$ for $\Omo = (0.20, 0.35)$. There are also exact analytic solutions of $D$ for general constant $\w$ dark energy models and thus one can obtain analytic form of $\gamma$ by using Eq. (\ref{f}) \cite{09061643,09072108}.

In general, $\gamma$ can be a function of time and Eq. (\ref{deltama}) can be rewritten by using $f$ and $\gamma$ \be f(a)^2 + \Biggl[\fr{1}{2} + 3\w \Bigl( 1-\Oma \Bigr) (\gamma-\fr{1}{2}) + \fr{d \gamma}{d \ln a} \ln \Oma \Biggr] f(a) = \fr{3}{2} \fr{G_{\eff}}{G_{\N}} \Oma \, . \label{fa} \ee  We adopt the currently available data of the growth function $f = \beta b$ derived from the redshift space distortion parameter $\beta(z)$ and the linear bias $b(z)$ in order to constrain the free parameter $\gamma$. However, the data referred as ``\obs'' in Table \ref{table1} have the different normalization ($\sigma_8$) and present matter density ($\Omo$). Thus, we adopt a rescaling method given in \cite{12021637} in order to transform different growth function to the same fiducial cosmology ({\it ie.}, flat $\Lambda$CDM with $\Omo = 0.273$ and $\sigma_8 = 0.811$) and denote the quantities obtained from this method with the subscription ``$\res$''. If one assumes that both the redshift space distortion parameter $\beta$ and the $\sigma_8$ are independent of fiducial model, then one can obtain the relation (see detail in \cite{12021637}) \be f_{\res}(z) = f_{\obs}(z) \fr{\sigma_{8, \obs}}{\sigma_8} \fr{D_{\obs}(z)}{D(z)} \, . \label{fobs} \ee

\begin{center}
    \begin{table}
    \begin{tabular}{ | c | c | c | c | c | c | c | c|}
    \hline
    $z$ & $f_{\obs}$ & $(\Omega_{\m0,\obs}, \sigma_{8,\obs})$ &  $\fr{G_{\eff}}{G_{\N}} |_{\obs}$ & $\fr{D_{\obs}}{D}$ & $f_{\res}$  & $\fr{G_{\eff}}{G_{\N}} |_{\res} $ & reference \\ \hline
    0.067 & 0.58 $\pm$ 0.11 & (0.25, 0.76) & $1.29^{+0.84}_{-0.71}$ & $\fr{0.72}{0.74}$ & $0.53 \pm 0.10$ & $1.05^{+0.74}_{-0.63}$ & \cite{12044725} \\ \hline
    0.15 & 0.49 $\pm$ 0.14 & (0.30, 0.90) & $0.42^{+0.79}_{-0.59}$ & $\fr{0.72}{0.71}$ & $0.55 \pm 0.16$ & $0.88^{+1.09}_{-0.84}$ & \cite{08021944} \\ \hline
    0.17 & 0.64 $\pm$ 0.12 & (0.30, 0.78) & $1.20^{+0.80}_{-0.68}$ & $\fr{0.71}{0.70}$ & $0.62 \pm 0.12$ & $1.26^{+0.85}_{-0.72}$ & \cite{08070810} \\ \hline
    0.35 & 0.70 $\pm$ 0.18 & (0.24, 0.76) & $1.40^{+1.23}_{-0.98}$ & $\fr{0.63}{0.64}$ & $0.64 \pm 0.17$ & $0.86^{+0.98}_{-0.77}$ & \cite{0608632} \\ \hline
    0.55$^{*}$ & 0.75 $\pm$ 0.18 & (0.3, 1.0) & $0.93^{+0.92}_{-0.74}$ & $\fr{0.59}{0.58}$ & $0.93 \pm 0.22$ & $2.01^{+1.41}_{-1.16}$ & \cite{0612400} \\ \hline
    0.77 & 0.91 $\pm$ 0.36 & (0.27, 0.78) & $1.51^{+2.09}_{-1.47}$ & $\fr{0.52}{0.52}$ & $0.87 \pm 0.35$ & $1.30^{+1.95}_{-1.36}$ & \cite{08021944} \\ \hline
    1.40 & 0.90 $\pm$ 0.24 & (0.25, 0.84) & $1.01^{+1.09}_{-0.85}$ & $\fr{0.40}{0.40}$ & $0.93 \pm 0.25$ & $1.09^{+1.14}_{-0.89}$ & \cite{0612401} \\ \hline
    2.42 & 0.74 $\pm$ 0.24 & (0.26, 0.93) & $0.25^{+0.82}_{-0.57}$ & $\fr{0.29}{0.29}$ & $0.85 \pm 0.28$ & $0.59^{+1.08}_{-0.78}$ & \cite{0404600} \\ \hline
    3.00$^{*}$ & 1.46 $\pm$ 0.29 & (0.30, 0.85) & $3.15^{+1.62}_{-1.39}$ & $\fr{0.25}{0.24}$ & $1.60 \pm 0.32$ & $3.93^{+1.93}_{-1.65}$ & \cite{0407377} \\ \hline
    0.22 & 0.60 $\pm$ 0.10 & (0.27, 0.80) & $0.99^{+0.63}_{-0.55}$ & $\fr{0.68}{0.68}$ & $0.59 \pm 0.10$ & $0.91^{+0.62}_{-0.53}$ & \cite{11042948} \\ \hline
    0.41 & 0.70 $\pm$ 0.07 & (0.27, 0.80) & $1.06^{+0.39}_{-0.35}$ & $\fr{0.62}{0.62}$ & $0.69 \pm 0.07$ & $0.99^{+0.38}_{-0.35}$ & \cite{11042948} \\ \hline
    0.44 & 0.64 $\pm$ 0.12 & (0.27, 0.80) & $0.70^{+0.61}_{-0.51}$ & $\fr{0.61}{0.61}$ & $0.63 \pm 0.12$ & $0.65^{+0.60}_{-0.50}$ & \cite{12043674} \\ \hline
    0.60 & 0.73 $\pm$ 0.07 & (0.27, 0.80) & $0.88^{+0.34}_{-0.31}$ & $\fr{0.57}{0.57}$ & $0.72 \pm 0.07$ & $0.82^{+0.33}_{-0.30}$ & \cite{11042948} \\ \hline
    0.73 & 0.78 $\pm$ 0.13 & (0.27, 0.80) & $0.94^{+0.63}_{-0.54}$ & $\fr{0.53}{0.53}$ & $0.77 \pm 0.13$ & $0.88^{+0.62}_{-0.53}$ & \cite{12043674} \\ \hline
    0.78 & 0.70 $\pm$ 0.08 & (0.27, 0.80) & $0.55^{+0.33}_{-0.30}$ & $\fr{0.50}{0.50}$ & $0.69 \pm 0.08$ & $0.51^{+0.33}_{-0.29}$ & \cite{11042948} \\ \hline
    \end{tabular}
    \caption{The currently available data for the growth functions $f_{\obs}$ and their re-scaled values $f_{\res}$. Also we show the derived values of $\Geff / \GN |_{\obs}$ and $\Geff / \GN |_{\res}$ from $f_{\obs}$ and $f_{\res}$, respectively.}
    \label{table1}
    \end{table}
\end{center}

In Table \ref{table1}, we show the currently available values of growth function $f_{\obs}(z)$ at the different redshifts which include 6 degree Field Galaxy Survey (6dFGS) at $z = 0.067$ \cite{12044725}, 2 degree Field Galaxy Redshift Survey (2dFGRS) at $z = 0.15$, VIMOS-VLT Deep Survey (VVDS) at $z = 0.77$ \cite{08021944}, 2dFGRS at $z = 0.17$ \cite{08070810}, Sloan Digital Sky Survey (SDSS) luminous red galaxies (LRGs) at $z = 0.35$, 2dF-SDSS LRG and Quasi Stellar Objects (QSO) Survey (2SLAQ) at $z = 0.55$ \cite{0612400}, 2dF QSO Redshift Survey (2QZ) and the 2dF-SDSS LRG and QSO Survey (2SLAQ) at $z = 1.40$ \cite{0612401}, Large Sample of UVES QSO Absorption Spectra (LUQAS) at $z = 2.42$ \cite{0404600}, Lyman-$\alpha$ forest in the SDSS at $z = 3.00$ \cite{0407377}, and WiggleZ Dark Energy Survey at $z = 0.22, 0.41, 0.44, 0.60, 0.73, 0.78$ \cite{11042948,12043674}. We also derive the ratio of the effective gravitational constant to the Newton's one from $f_{\obs}$ and represent them as $\Geff / \GN |_{\obs}$. We also list the rescaled values of $f_{\res}$ and $\Geff / \GN |_{\res}$.

We constraint parametrization of $\gamma(z)$ with the data given in Table \ref{table1}. We use phenomenological parametrization of $\gamma(z)$ by using the general relation \be \gamma(z) = \gamma_{0} + \gamma_{a} p(z) \, . \label{gammaz} \ee We define Model 1 \cite{07101510}, 2 \cite{10043086}, 3, and 4  with $p(z) = (z, \fr{z}{1+z}, \fr{1}{1+z})$, and $0$, respectively. Model 3 is also investigated, because $\gamma_a$ is proportional to $a$ in the constant $\w$ models \cite{09051522}.

We perform a standard $\chi^2$ minimization with two quantities $f$ and $g \equiv G_{\eff}/G_{\N}$ to constrain the model parameters $\gamma_{0}$ and $\gamma_{a}$. \be \chi^2(z_i | \vec{p}) = \sum_{i=1}^{N} \Biggl( \fr{{\cal O}_{\obs}(z_i) - {\cal O}_{\rm{theor}}(z_i, \vec{p})}{\sigma_i} \Biggr)^2 \, , \label{chi2} \ee where $N$ is the number of observations, ${\cal O}_{\obs (\rm{theor})}$ is the observational (theoretical) values of observations, $\vec{p}$ is the fitted parameters, and $\sigma_i$ is the uncertainties of observations. We repeat $\chi^2$ minimization by using ``observed'' and ``rescaled'' values separately for different models.

\begin{figure}
\centering
\vspace{1.5cm}
\begin{tabular}{ccc}
\epsfig{file=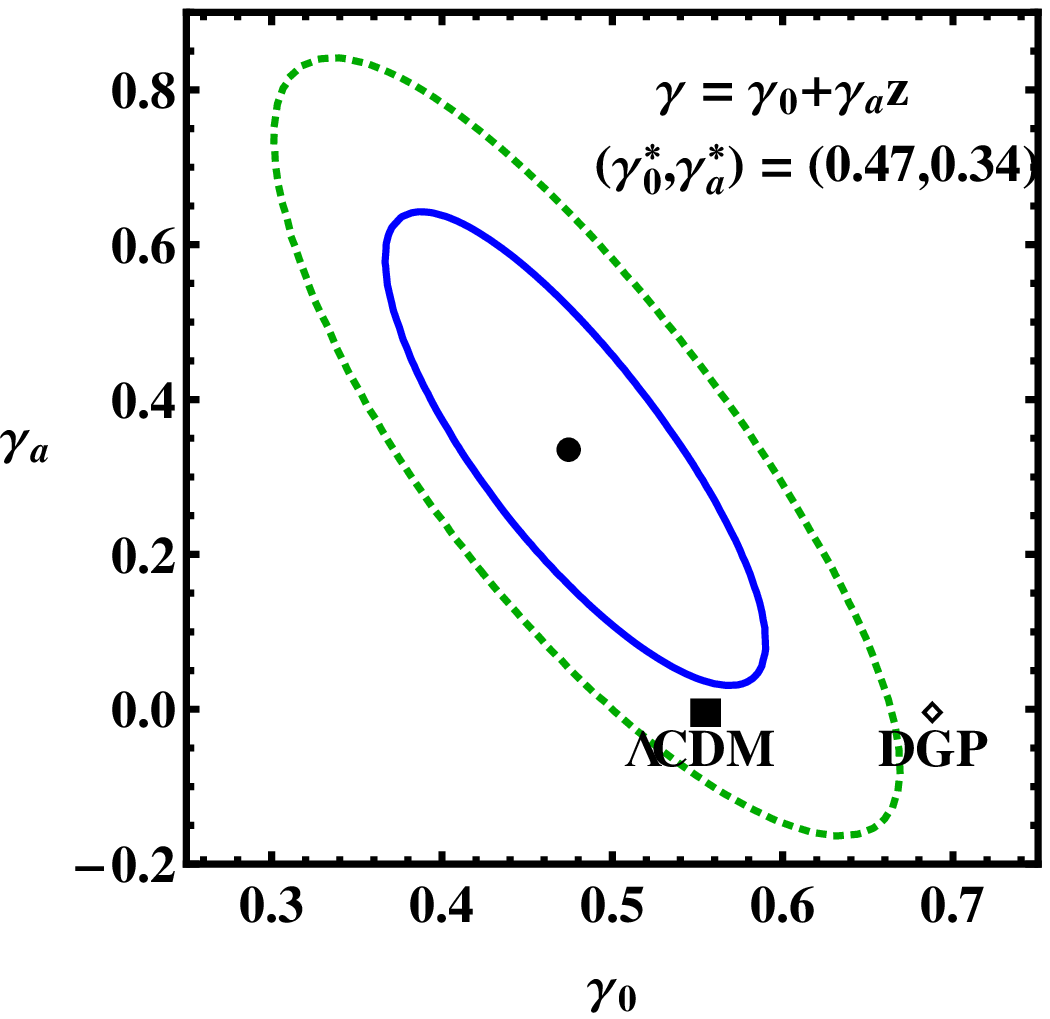,width=0.3\linewidth,clip=} &
\epsfig{file=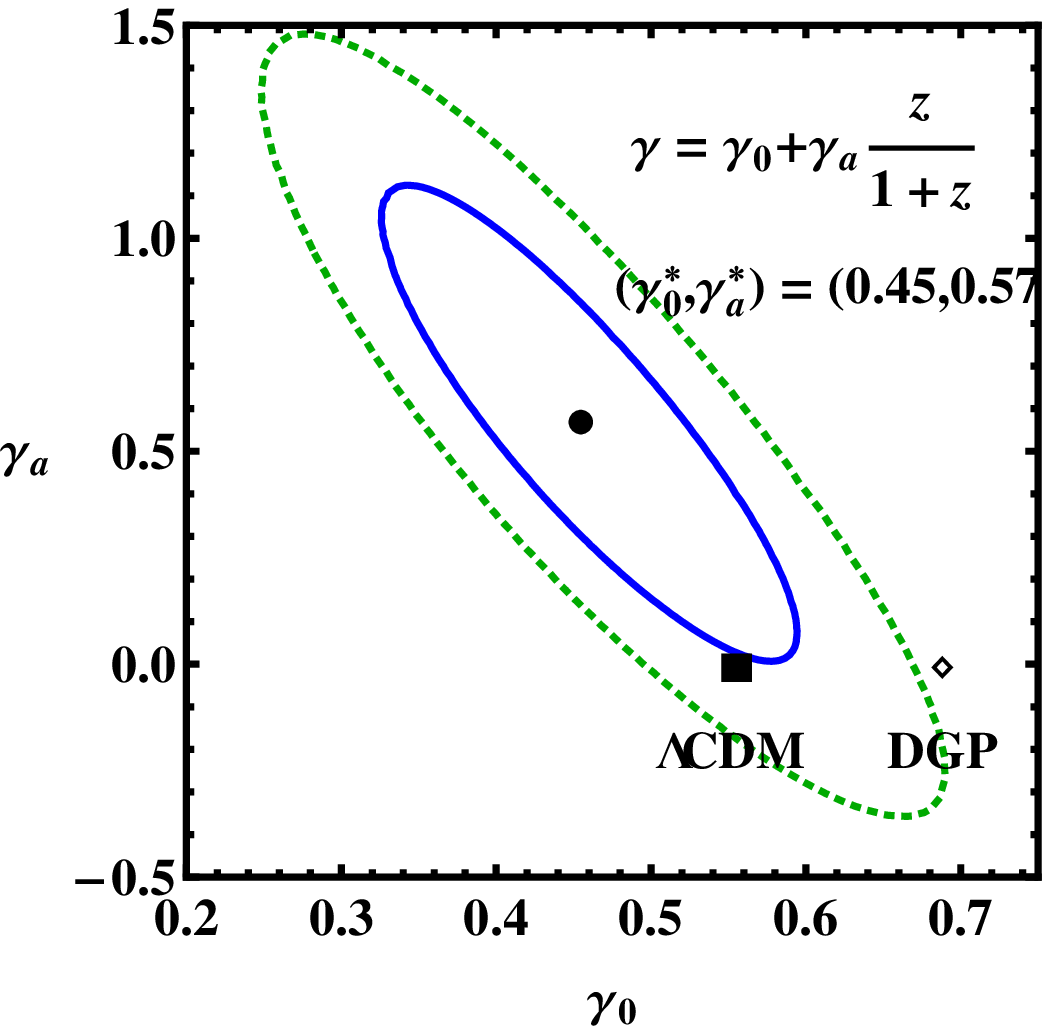,width=0.3\linewidth,clip=} &
\epsfig{file=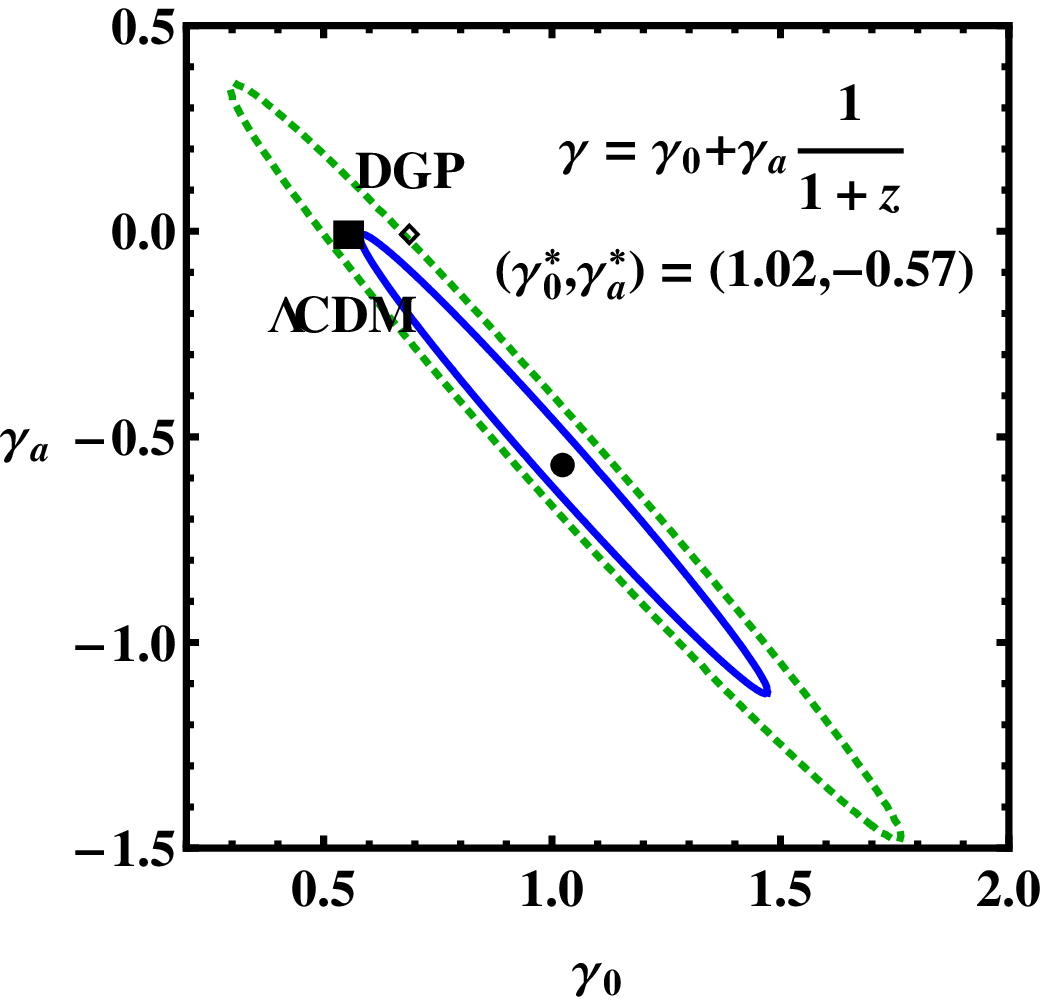,width=0.3\linewidth,clip=} \\
\epsfig{file=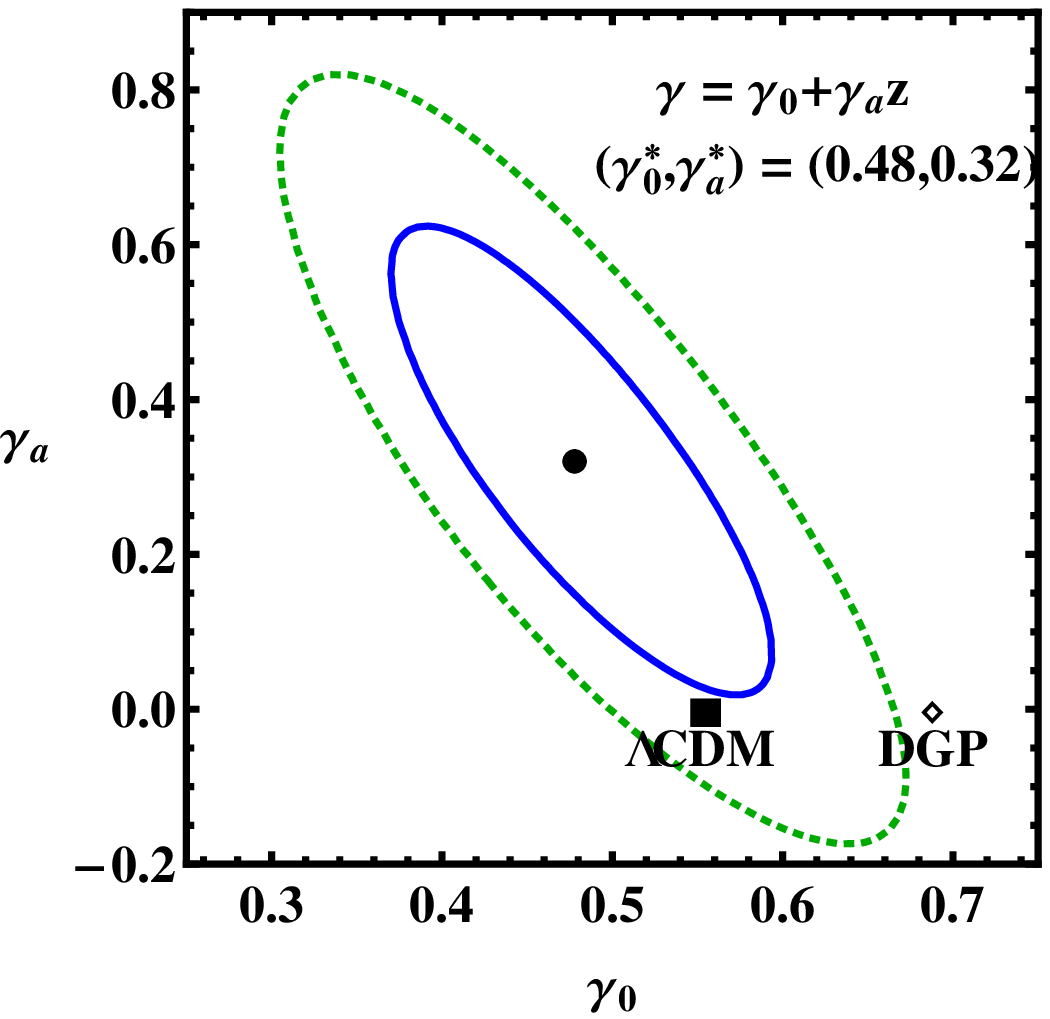,width=0.3\linewidth,clip=} &
\epsfig{file=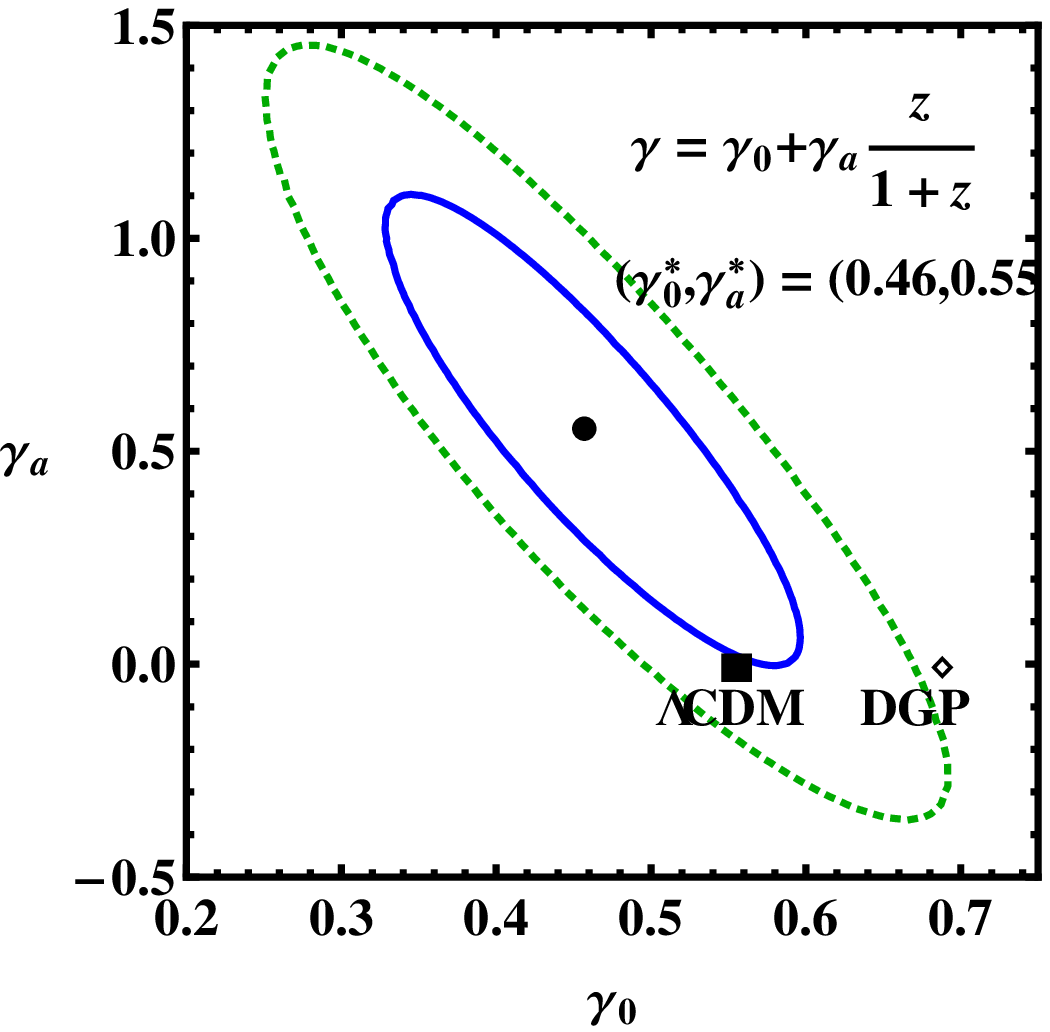,width=0.3\linewidth,clip=} &
\epsfig{file=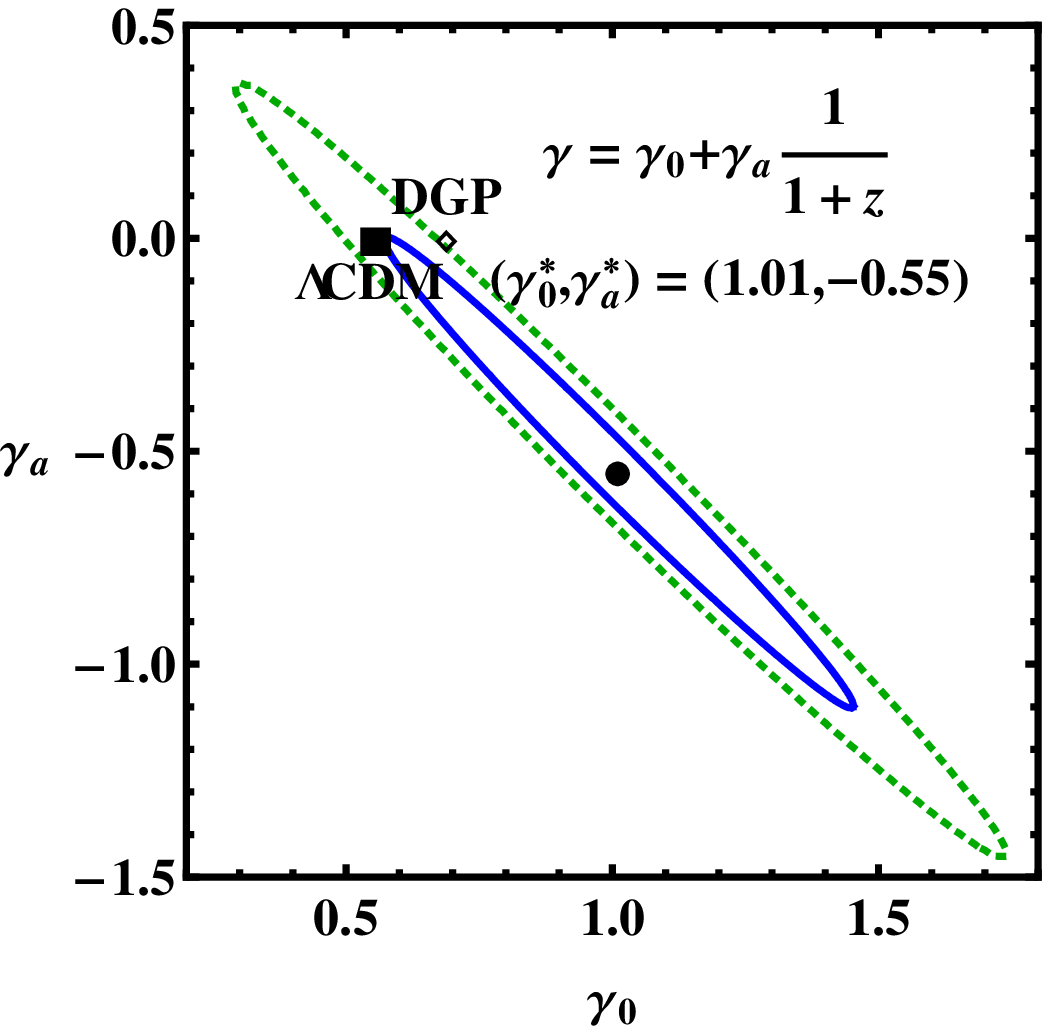,width=0.3\linewidth,clip=} \\
\end{tabular}
\vspace{-0.5cm}
\caption{ a) Likelihood contours for 1-$\sigma$ (solid lines) and 2-$\sigma$ (dotted lines) confidence levels of the ($\gamma_{0}, \gamma_a$) plane in the case of Model 1 (first column), Model 2 (second column), and Model 3 (third column). The first row is obtained with using Data 1. b) Same contours for the same models by using Data 2.} \label{fig1}
\end{figure}

First, we constrain the $\gamma$ by using $f_{\obs}$ with 13 (without asterisk) and 15 data points from Table \ref{table1}. We separate the two data points from others with asterisk, because the data obtained from the Lyman-$\alpha$ forest at $z = 3.0$ gives $f > 1$ which is physically unreasonable. Also, 2SLAQ data at $z =0.55$ produces too large $G_{eff}/G_{N} |_{\obs}$. We define Data 1 and Data 2 for 13 and 15 data points, respectively. We define the reduced $\chi^2$ as $\chi^2_{\rm{red}} = \chi^2 / dof$ with degree of freedom $dof = (N - n -1)$ where $n$ is the number of fitted parameters. The results for $\chi^2$ minimization of $f_{\obs}$ are shown in Table \ref{table2} and Fig. \ref{fig1}. M1 provides $(\gamma_0, \gamma_a) = (0.47 \pm 0.07, 0.34 \pm 0.20)$ with $\chi_{\rm{min}}^2 /dof = 7.42/10$ at 1-$\sigma$ level for Data 1 and $(\gamma_0, \gamma_a) = (0.48 \pm 0.07, 0.32 \pm 0.20)$ with $\chi_{\rm{min}}^2 /dof = 8.91/12$ for Data 2, respectively. Data 1 gives the almost same good fit as Data 2. For M2, $(\gamma_0, \gamma_a) = (0.45 \pm 0.09, 0.57 \pm 0.37)$ with $\chi_{\rm{min}}^2 /dof = 7.85/10$ and $(\gamma_0, \gamma_a) = (0.46 \pm 0.09, 0.55 \pm 0.36)$ with $\chi_{\rm{min}}^2 /dof = 9.22/12$ for Data 1 and 2, respectively. If we use M3, then $(\gamma_0, \gamma_a) = (1.02 \pm 0.29, -0.57 \pm 0.37)$ with $\chi_{\rm{min}}^2 /dof = 7.85/10$ and $(\gamma_0, \gamma_a) = (1.01 \pm 0.29, -0.55 \pm 0.36)$ with $\chi_{\rm{min}}^2 /dof = 9.23/12$ for Data 1 and 2, respectively. M4 gives $\gamma_0 = 0.58 \pm 0.04$ with $\chi_{\rm{min}}^2 /dof = 10.21/10$ and $\gamma_0 = 0.58 \pm 0.04$ with $\chi_{\rm{min}}^2 /dof = 11.50/12$ for Data 1 and 2, respectively. Thus, M1 is the best fit model and M4 is the worst one. M3 shows the almost same good fit as M2 but with large errors in $\gamma_0$. In Fig. \ref{fig1}, we show the 1 and 2-$\sigma$ likelihood contours of ($\gamma_0, \gamma_a$) plane for M1, M2, and M3 with Data 1 in the first row. We also show the same contours plots of same models with Data 2 in the second row. As shown in Fig. \ref{fig1}, $\gamma$ values of all models exclude the $\Lambda$CDM expected value of $\gamma(\Omo=0.273) = 0.555$ at 1-$\sigma$ level except M4. Also one of well known large extra dimension models of Dvali, Gabadadze, and Porrati (DGP) with $\gamma = 11/16$ is excluded for all models at 2-$\sigma$ level. However, $\gamma$ value of DGP is obtained from the different background evolution ({\it i.e} with $\w$ different from -1) and thus one is not able to conclude for the validity of DGP model with data in Table \ref{table1}.

\begin{center}
    \begin{table}
    \begin{tabular}{ | c | c | c | c | c | c | c | c | c | }
    \hline
    f & \multicolumn{4}{|c|}{Data 1} & \multicolumn{4}{|c|}{Data 2} \\ \cline{2-9}
      & $\gamma_{0}$ & $\gamma_{a}$ & $\chi_{\rm{min}}^2$ & $\chi_{\rm{red},\rm{min}}^2$ & $\gamma_{0}$ & $\gamma_{a}$ & $\chi_{\rm{min}}^2$ & $\chi_{\rm{red},\rm{min}}^2$ \\ \hline
    M1 & 0.47 $\pm$ 0.07 & 0.34 $\pm$ 0.20 & 7.42 & 0.74 & 0.48 $\pm$ 0.07 & 0.32 $\pm$ 0.20 & 8.91 & 0.74 \\ \hline
    M2 & 0.45 $\pm$ 0.09 & 0.57 $\pm$ 0.37 & 7.85 & 0.79 & 0.46 $\pm$ 0.09 & 0.55 $\pm$ 0.36 & 9.22 & 0.77 \\ \hline
    M3 & 1.02 $\pm$ 0.29 & -0.57 $\pm$ 0.37& 7.85 & 0.79 & 1.01 $\pm$ 0.29 & -0.55 $\pm$ 0.36 & 9.23& 0.77 \\ \hline
    M4 & 0.58 $\pm$ 0.04 & 0               & 10.21& 1.02 & 0.58 $\pm$ 0.04 & 0               &11.50 & 0.88 \\ \hline
    \end{tabular}
    \caption{$\gamma_{0}$ and $\gamma_{a}$ obtained from $\chi^2$ minimization of $f_{\obs}$ for the different models with Data 1 and 2. }
    \label{table2}
    \end{table}
\end{center}

\begin{figure}
\centering
\vspace{1.5cm}
\begin{tabular}{ccc}
\epsfig{file=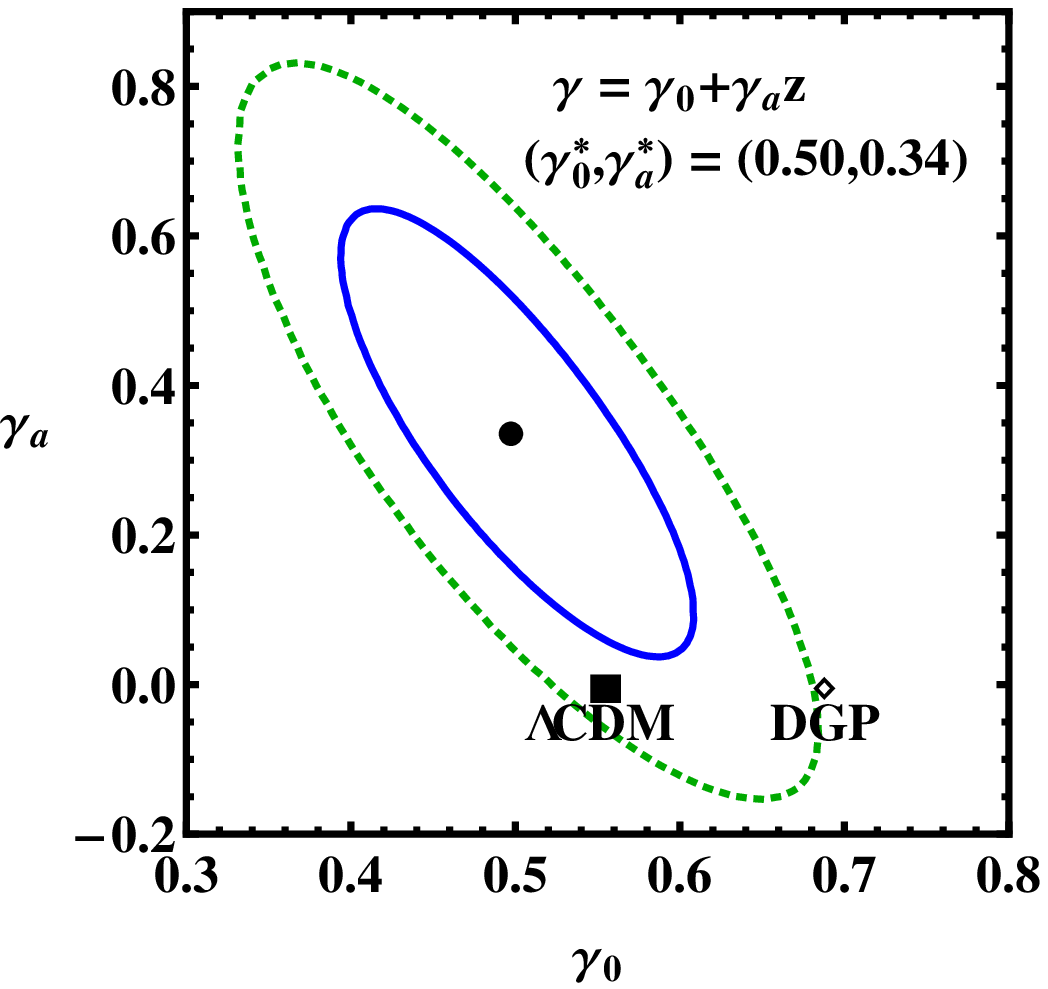,width=0.3\linewidth,clip=} &
\epsfig{file=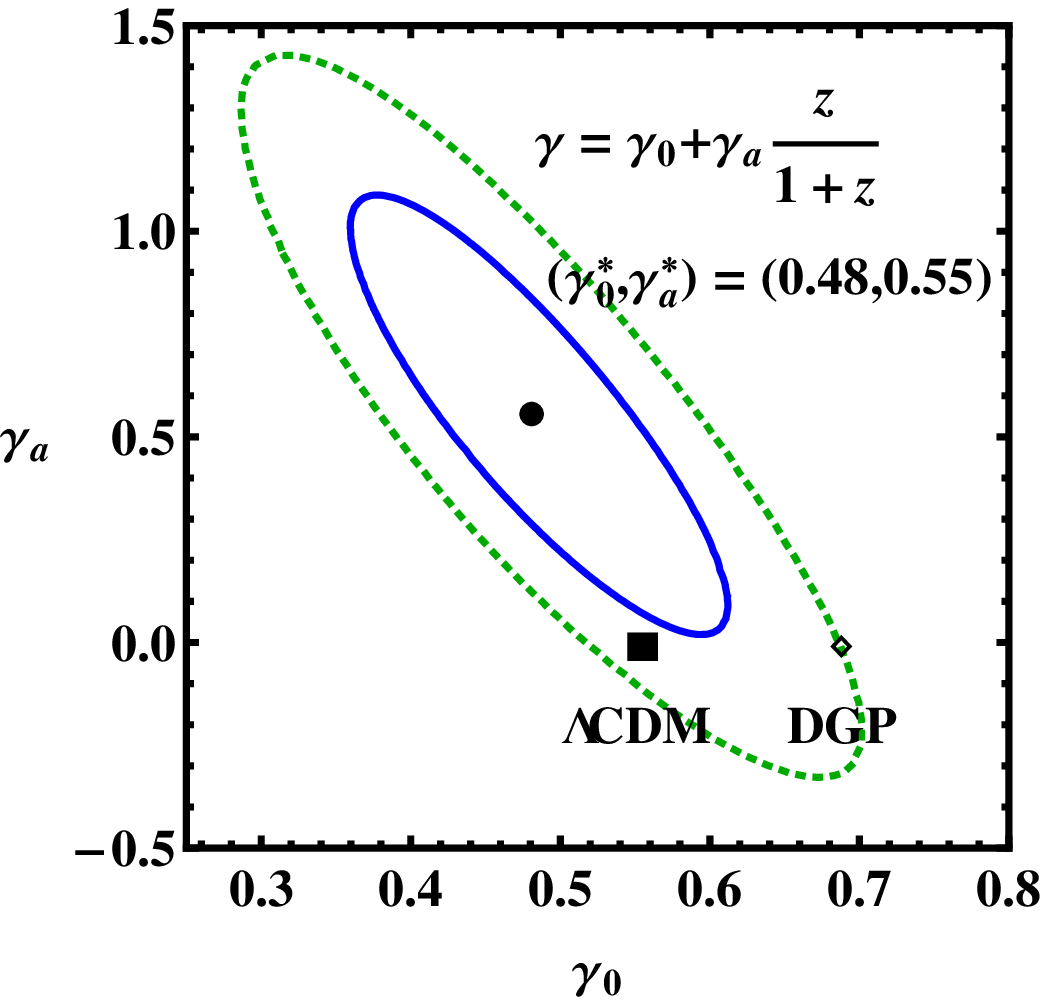,width=0.3\linewidth,clip=} &
\epsfig{file=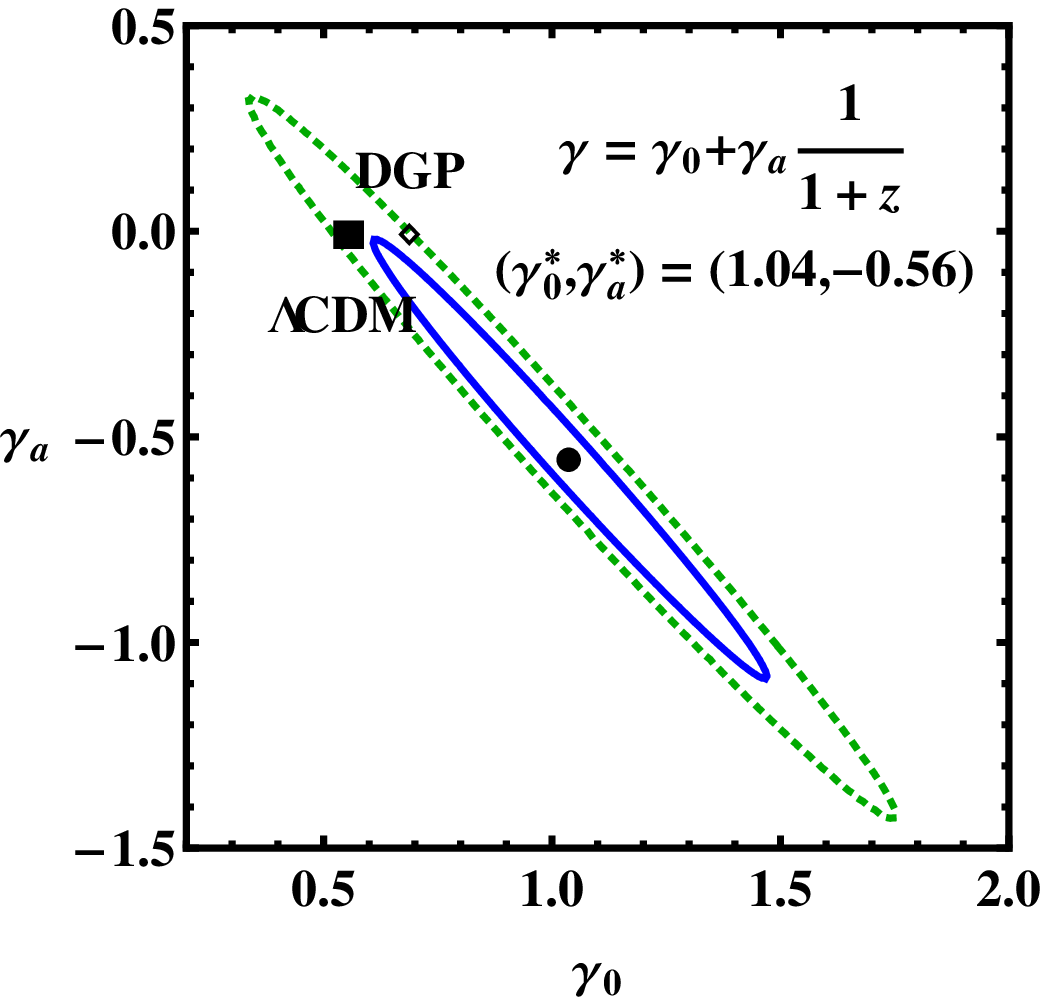,width=0.3\linewidth,clip=} \\
\epsfig{file=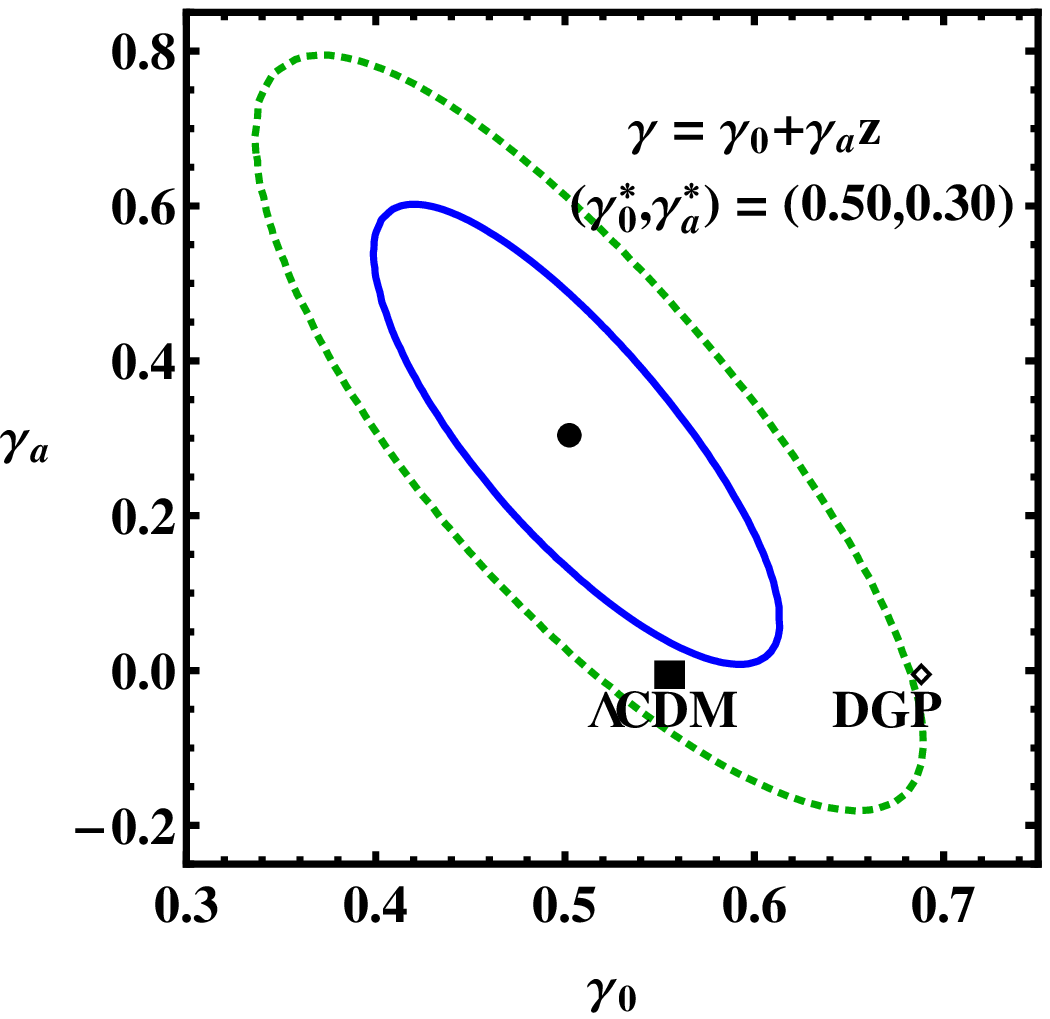,width=0.3\linewidth,clip=} &
\epsfig{file=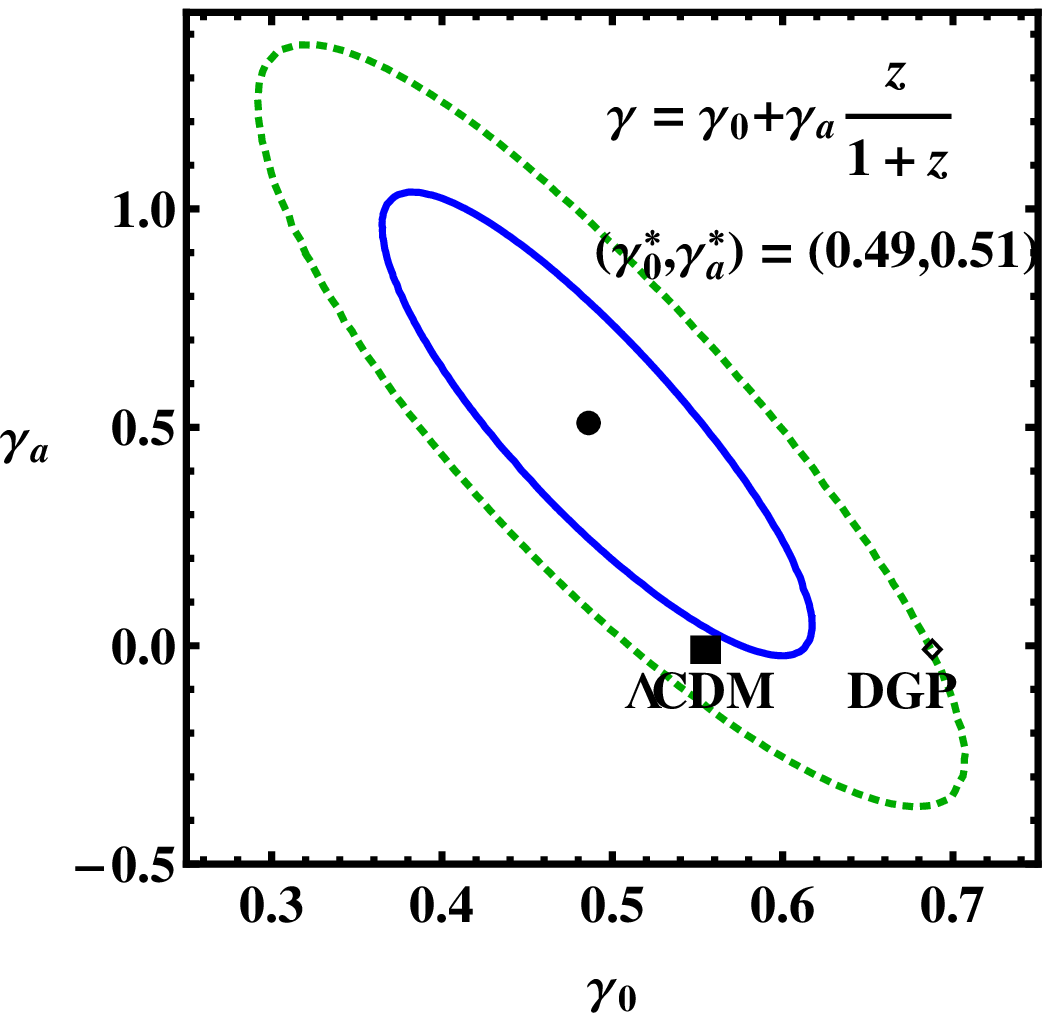,width=0.3\linewidth,clip=} &
\epsfig{file=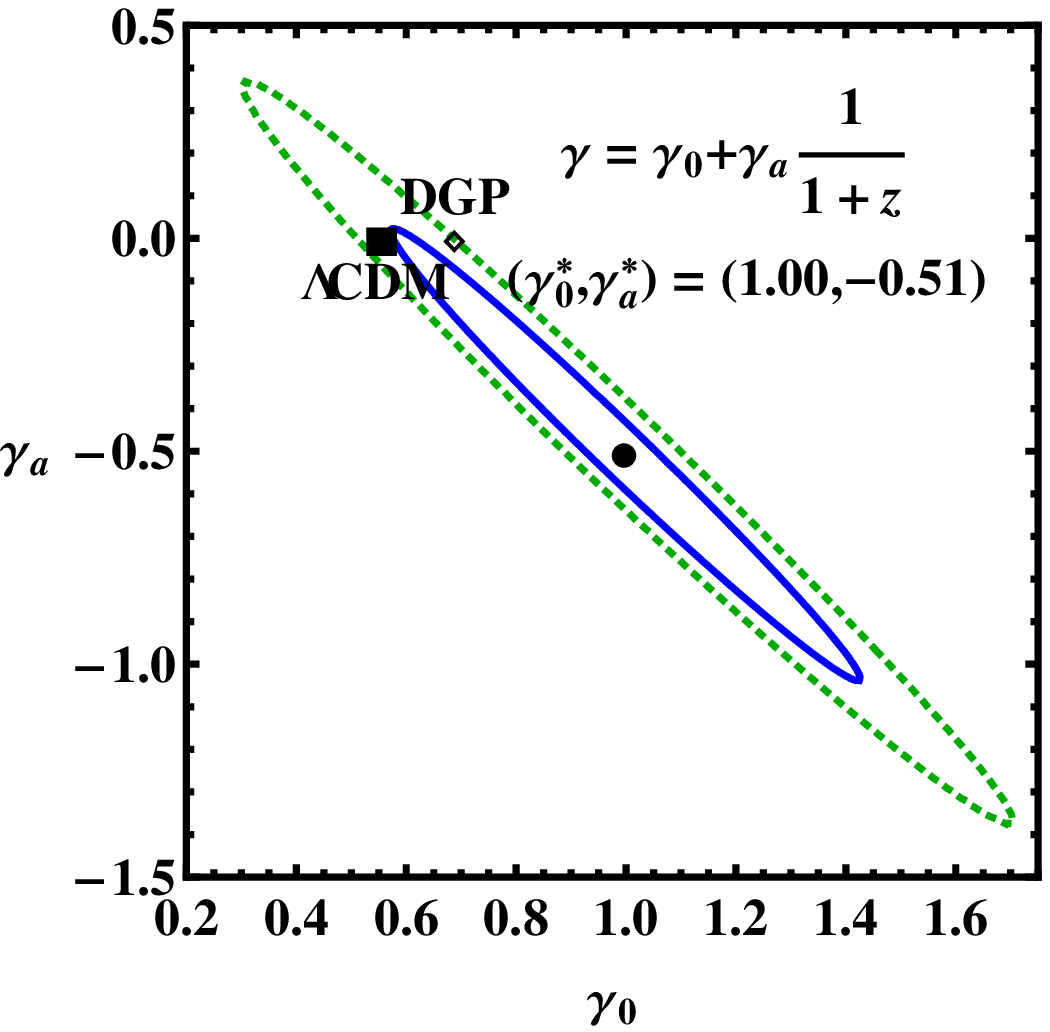,width=0.3\linewidth,clip=} \\
\end{tabular}
\vspace{-0.5cm}
\caption{ a) Likelihood contours for 1-$\sigma$ (solid lines) and 2-$\sigma$ (dotted lines) confidence levels of the ($\gamma_{0}, \gamma_a$) plane in the case of Model 1 (first column), Model 2 (second column), and Model 3 (third column). The first row is obtained with using Data 3. b) Same contours for the same models by using Data 4.} \label{fig2}
\end{figure}

Now, we repeat the constraint of $\gamma$ by using $f_{\res}$ with 13 (without asterisk) and 15 data points from Table \ref{table1}. We define Data 3 and Data 4 for 13 and 15 data points, respectively. The results for $\chi^2$ minimization of $f_{\res}$ are shown in Table \ref{table3} and Fig. \ref{fig2}. M1 provides $(\gamma_0, \gamma_a) = (0.50 \pm 0.07, 0.34 \pm 0.20)$ at 1-$\sigma$ level with $\chi_{\rm{min}}^2 /dof = 2.83/10$ for Data 3 and $(\gamma_0, \gamma_a) = (0.50 \pm 0.07, 0.30 \pm 0.20)$ with $\chi_{\rm{min}}^2 /dof = 5.80/12$ for Data 4, respectively. As expected Data 3 gives the better fit than Data 4. For M2, $(\gamma_0, \gamma_a) = (0.48 \pm 0.08, 0.56 \pm 0.35)$ with $\chi_{\rm{min}}^2 /dof = 3.27/10$ and $(\gamma_0, \gamma_a) = (0.49 \pm 0.08, 0.51 \pm 0.35)$ with $\chi_{\rm{min}}^2 /dof = 6.12/12$ for Data 3 and 4, respectively. If we use M3, then $(\gamma_0, \gamma_a) = (1.04 \pm 0.28, -0.56 \pm 0.35)$ with $\chi_{\rm{min}}^2 /dof = 3.27/10$ and $(\gamma_0, \gamma_a) = (1.00 \pm 0.28, -0.51 \pm 0.35)$ with $\chi_{\rm{min}}^2 /dof = 6.12/12$ for Data 3 and 4, respectively. M4 gives $\gamma_0 = 0.60 \pm 0.04$ with $\chi_{\rm{min}}^2 /dof = 5.73/10$ and $\gamma_0 = 0.59 \pm 0.04$ with $\chi_{\rm{min}}^2 /dof = 8.22/12$ for Data 3 and 4, respectively. Again M1 is the best fit model and M4 is the worst one. In Fig. \ref{fig2}, we show the 1 and 2-$\sigma$ likelihood contours of ($\gamma_0, \gamma_a$) plane for M1, M2, and M3 with Data 3 in the first row. We also show the same contours plots of same models with Data 4 in the second row. As shown in Fig. \ref{fig2}, $\gamma$ values of all models exclude the $\Lambda$CDM expected value of $\gamma(\Omo=0.273) = 0.555$ at 1-$\sigma$ level except M4. Also DGP with $\gamma = 11/16$ is marginally excluded for all models at 2-$\sigma$ level. Compared to the constraints from Data 1 and 2, $f_{\res}$ give the better fits for all models. The above results are quite similar to the recent analysis by using $f \sigma_8$ data \cite{12036724}.

\begin{center}
    \begin{table}
    \begin{tabular}{ | c | c | c | c | c | c | c | c | c | }
    \hline
    f & \multicolumn{4}{|c|}{Data 3} & \multicolumn{4}{|c|}{Data 4} \\ \cline{2-9}
      & $\gamma_{0}$ & $\gamma_{a}$ & $\chi_{\rm{min}}^2$ & $\chi_{\rm{red},\rm{min}}^2$ & $\gamma_{0}$ & $\gamma_{a}$ & $\chi_{\rm{min}}^2$ & $\chi_{\rm{red},\rm{min}}^2$ \\ \hline
    M1 & 0.50 $\pm$ 0.07 & 0.34 $\pm$ 0.20 & 2.83 & 0.28 & 0.50 $\pm$ 0.07 & 0.30 $\pm$ 0.20 & 5.80 & 0.48 \\ \hline
    M2 & 0.48 $\pm$ 0.08 & 0.56 $\pm$ 0.35 & 3.27 & 0.33 & 0.49 $\pm$ 0.08 & 0.51 $\pm$ 0.35 & 6.12 & 0.51 \\ \hline
    M3 & 1.04 $\pm$ 0.28 & -0.56 $\pm$ 0.35& 3.27 & 0.33 & 1.00 $\pm$ 0.28 & -0.51 $\pm$ 0.35 & 6.12& 0.51 \\ \hline
    M4 & 0.60 $\pm$ 0.04 & 0               & 5.73 & 0.52 & 0.59 $\pm$ 0.04 & 0               & 8.22 & 0.63 \\ \hline
    \end{tabular}
    \caption{$\gamma_{0}$ and $\gamma_{a}$ obtained from $\chi^2$ minimization of $f_{\res}$ for the different models by using Data 3 and 4. }
    \label{table3}
    \end{table}
\end{center}

\section{Constraints on Time varying $G$}
\setcounter{equation}{0}

Equation (\ref{fa}) can also be used to check any time variation of the effective gravitational constant \be g(a) \equiv \fr{G_{\eff}}{G_{\N}} = \Biggl( f(a)^2 + \Bigl[\fr{1}{2} + 3\w \Bigl( 1-\Oma \Bigr) (\gamma-\fr{1}{2}) + \fr{d \gamma}{d \ln a} \ln \Oma \Bigr] f(a) \Biggr) \fr{2}{3 \Oma} \, , \label{g} \ee where $\gamma$ can be obtained from the data by using its definition $\gamma = \ln[f_{\obs (\res)}] / \ln[\Omega_{\m, \obs (\res)}]$ for the observed (rescaled) quantities . Thus, $\fr{\Geff}{\GN}$ is obtained from Eq. (\ref{g}) by adopting the fiducial $\Lambda$CDM model where we assume that $\gamma$ is a constant. The observed (rescaled) values of $g_{\obs (\res)}$ are listed in Table \ref{table1}. We repeat the $\chi^2$ minimization to constrain the $\gamma$ parameters by using data from $g$. We separate Data 1 (3) and 2 (4) with and without $g_{\obs (\res)}$ values at $z = 0.55$ and $3.00$ again.

\begin{figure}
\centering
\vspace{1.5cm}
\begin{tabular}{ccc}
\epsfig{file=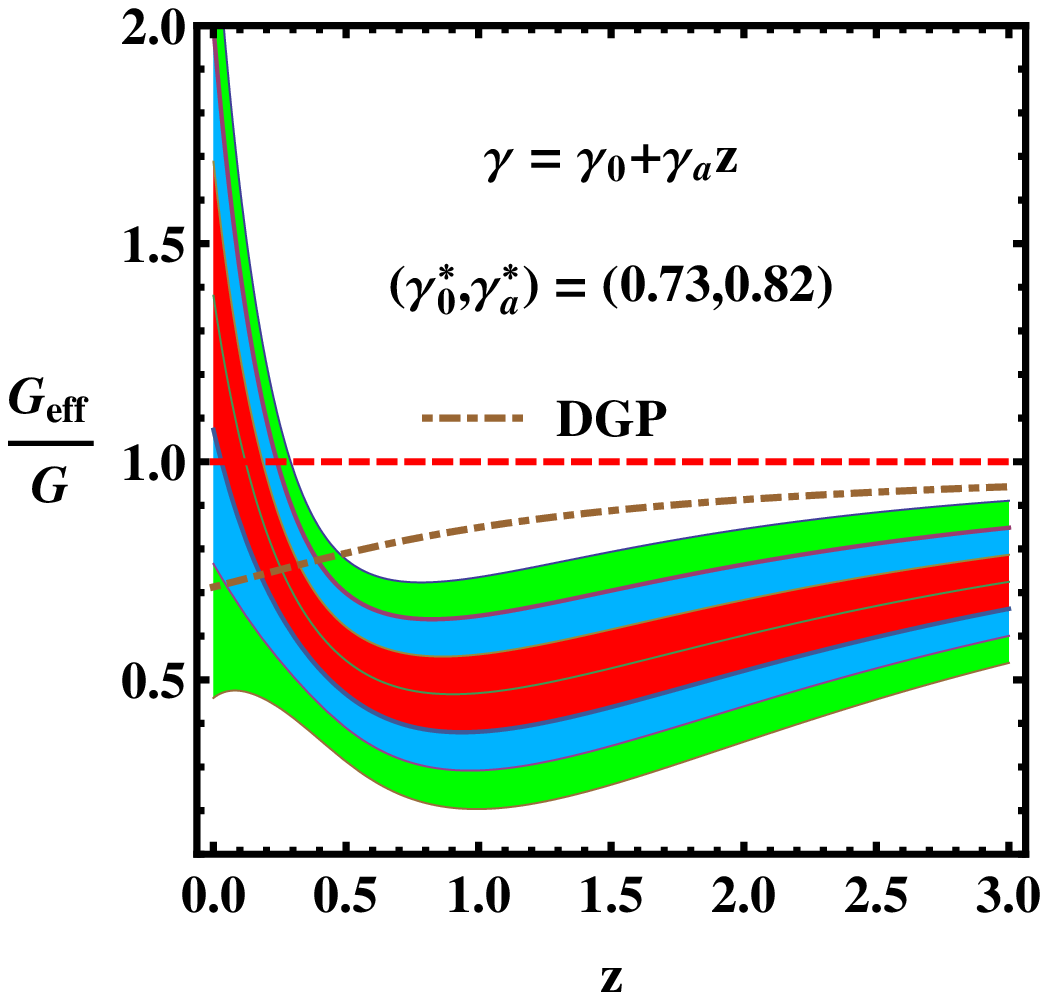,width=0.3\linewidth,clip=} &
\epsfig{file=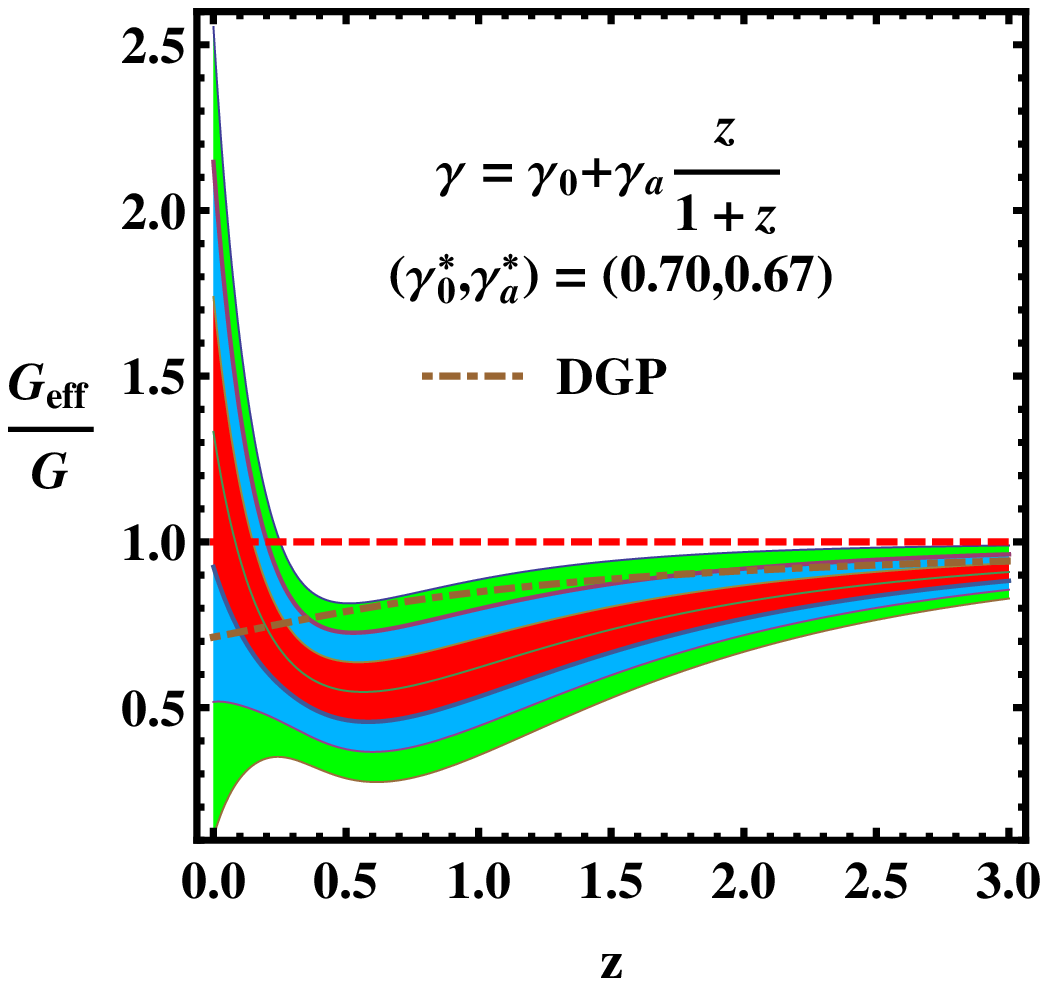,width=0.3\linewidth,clip=} &
\epsfig{file=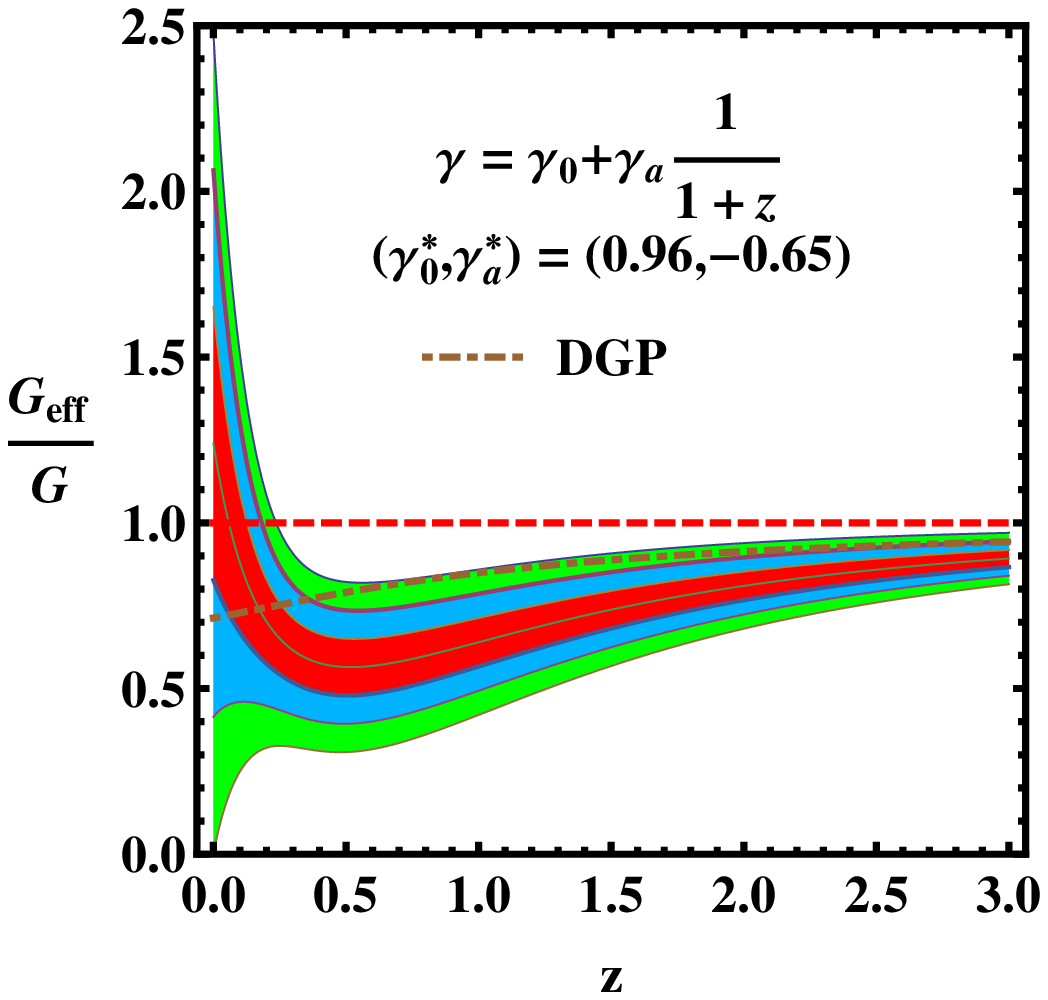,width=0.3\linewidth,clip=} \\
\epsfig{file=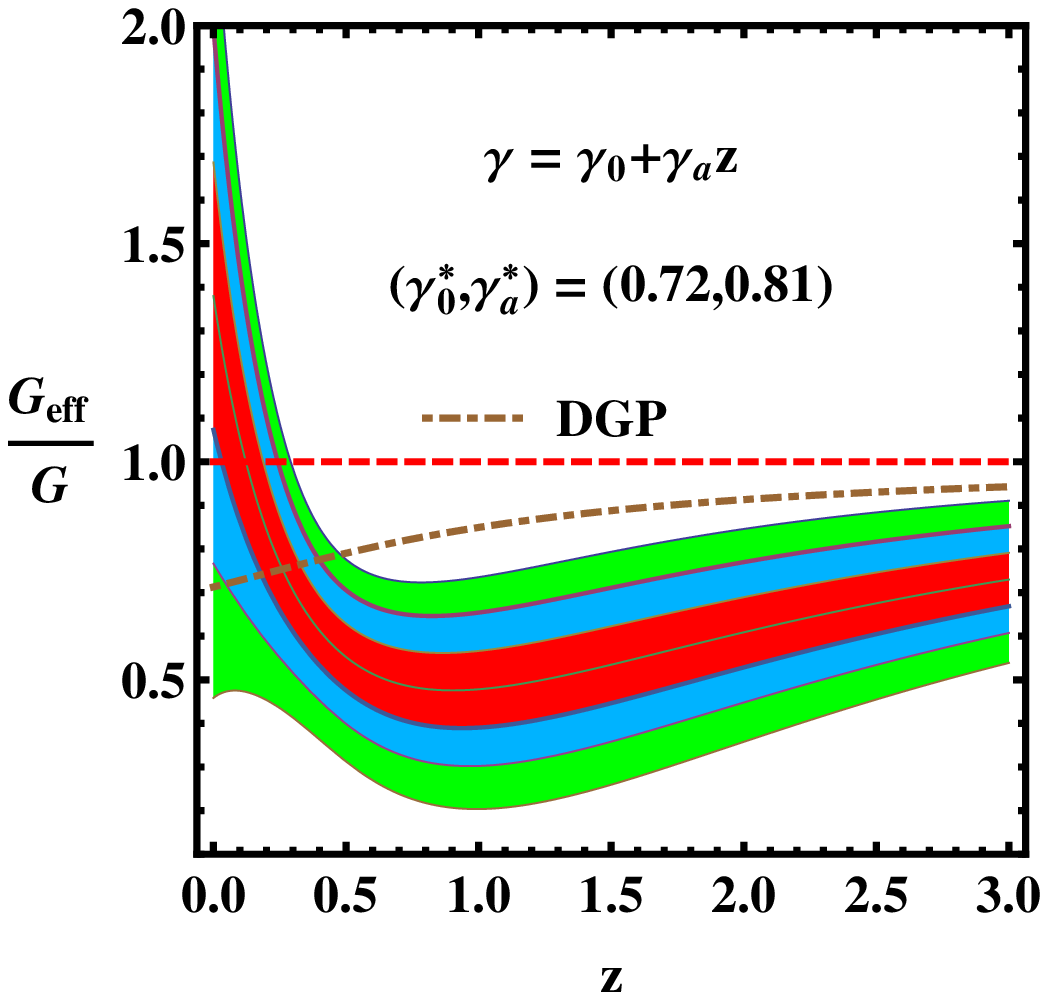,width=0.3\linewidth,clip=} &
\epsfig{file=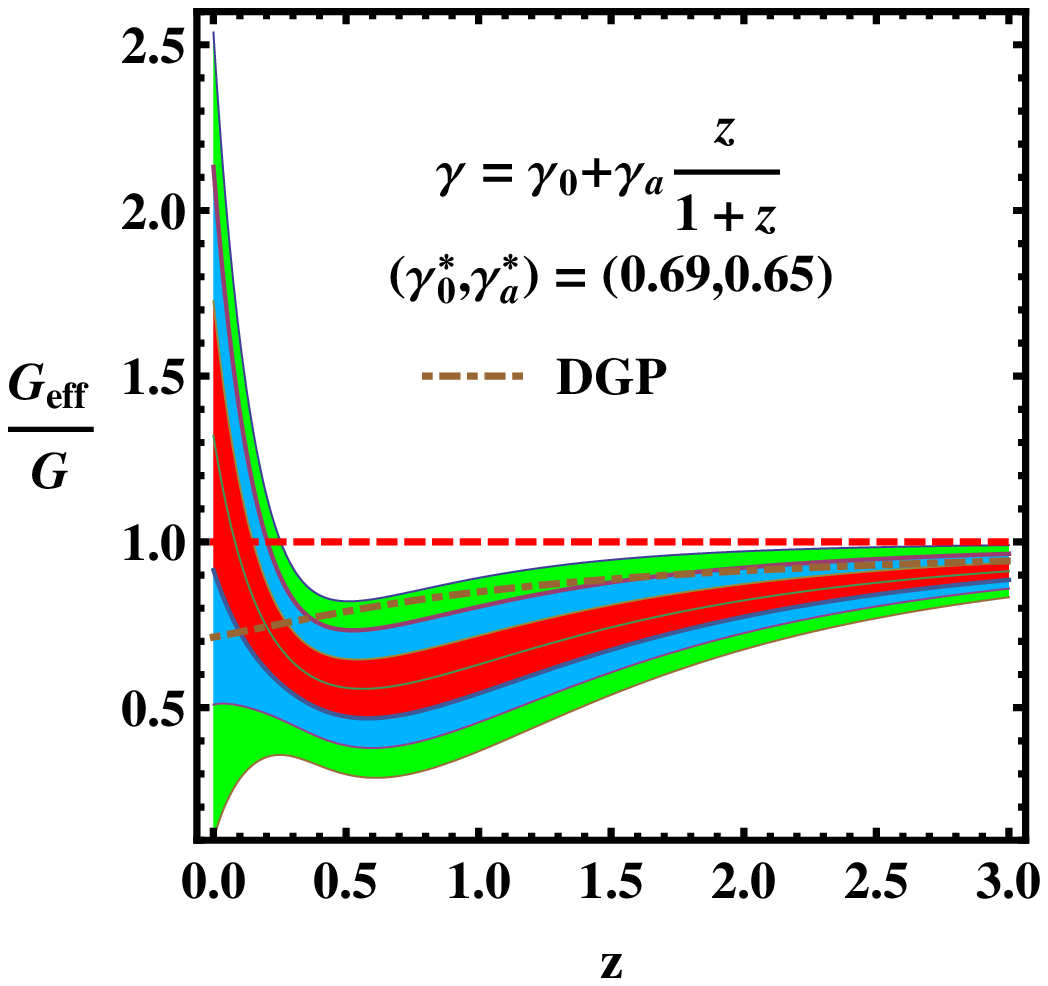,width=0.3\linewidth,clip=} &
\epsfig{file=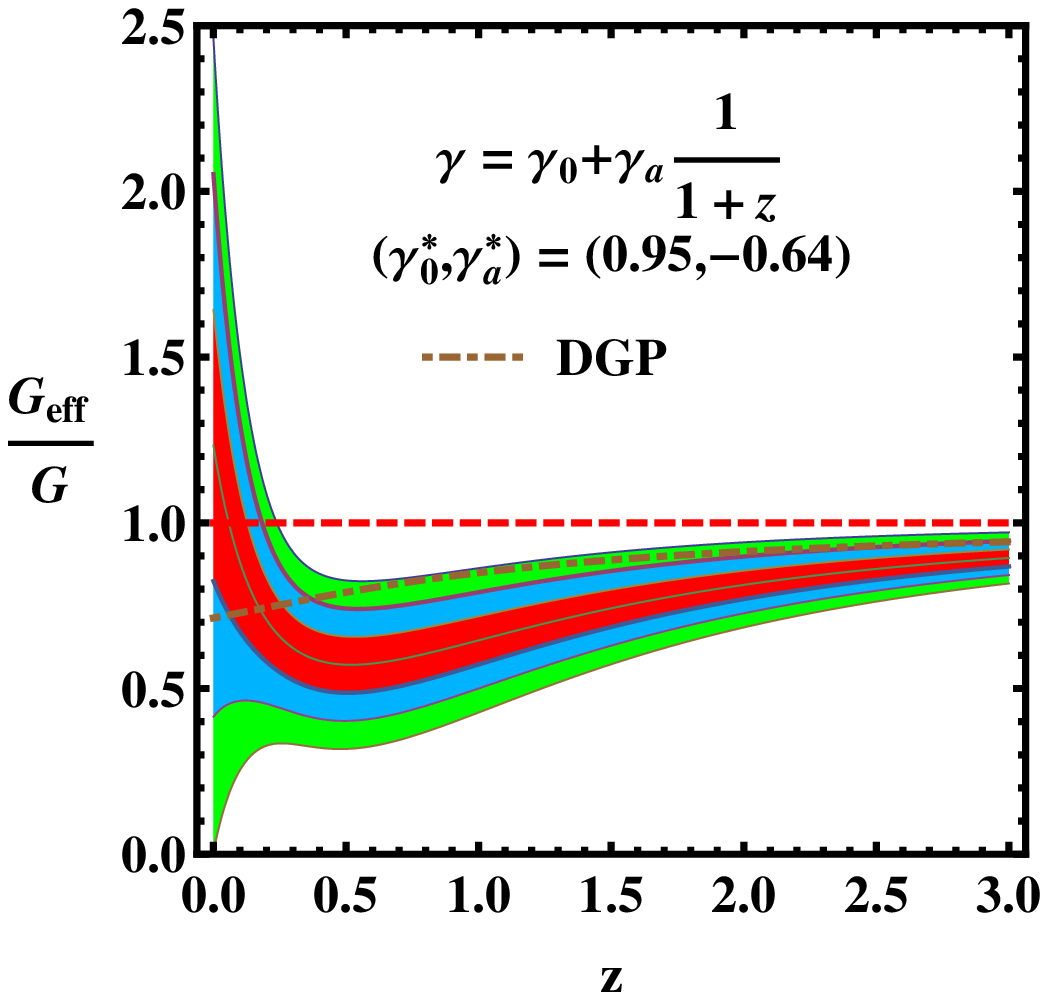,width=0.3\linewidth,clip=} \\
\end{tabular}
\vspace{-0.5cm}
\caption{ a) The best fit form of $\Geff / \GN |_{\obs}$ for 1-$\sigma$ (darker shaded regions), 2-$\sigma$ (dark shaded regions), and 3-$\sigma$ (light shaded regions) errors in the case of Model 1 (first column), Model 2 (second column), and Model 3 (third column). The first row is obtained with using Data 1. b) Same best fit form of $\Geff / \GN$ the same models with using Data 2.} \label{fig3}
\end{figure}

We show the results for $\chi^2$ minimization of $g_{\obs}$ in Table \ref{table4} and Fig. \ref{fig3}. M1 provides $(\gamma_0, \gamma_a) = (0.73 \pm 0.05, 0.82 \pm 0.22)$ at 1-$\sigma$ level with $\chi_{\rm{min}}^2 /dof = 27.65/10$ for Data 1 and $(\gamma_0, \gamma_a) = (0.72 \pm 0.05, 0.81 \pm 0.22)$ with $\chi_{\rm{min}}^2 /dof = 28.67/12$ for Data 2, respectively. Data 2 gives the better fit than Data 1. When we use M2, $(\gamma_0, \gamma_a) = (0.70 \pm 0.05, 0.67 \pm 0.26)$ with $\chi_{\rm{min}}^2 /dof = 40.38/10$ and $(\gamma_0, \gamma_a) = (0.69 \pm 0.05, 0.65 \pm 0.26)$ with $\chi_{\rm{min}}^2 /dof = 41.25/12$ for Data 1 Dnd 2, respectively. For M3, $(\gamma_0, \gamma_a) = (0.96 \pm 0.09, -0.65 \pm 0.19)$ with $\chi_{\rm{min}}^2 /dof = 39.09/10$ and $(\gamma_0, \gamma_a) = (0.95 \pm 0.09, -0.64 \pm 0.19)$ with $\chi_{\rm{min}}^2 /dof = 39.88/12$ for Data 1 and 2, respectively. From M4, we obtain $\gamma_0 = 0.72 \pm 0.05$ with $\chi_{\rm{min}}^2 /dof = 48.13/10$ and $\gamma_0 = 0.72 \pm 0.05$ with $\chi_{\rm{min}}^2 /dof = 48.70/12$ for Data 1 and 2, respectively. Again, M1 is the best fit and M4 is the worst one. In Fig. \ref{fig3}, we show the 1-$\sigma$ (inner darkest shaded regions), 2-$\sigma$ (dark shaded regions) and 3-$\sigma$ (light shaded ones) best fit form contours of $g(z)$ for M1, M2, and M3 with Data 1 in the first row. We also show the same contours plots of same models with Data 2 in the second row. Interestingly, $G_{\eff}/G_{N}$ deviates from $\Lambda$CDM expected value 1 from $z \gtrsim 0.2 \sim 0.3$ for all models at 3-$\sigma$ level. DGP expected value of $g(z)$ is also shown as the dot-dashed line. However, $g(z)$ of DGP is obtained from the effective equation of state of dark energy component $\w = -1 /(1+ \Omz)$ which is different from the fiducial model and thus it is impossible to investigate the viability of DGP model with this data.

\begin{center}
    \begin{table}
    \begin{tabular}{ | c | c | c | c | c | c | c | c | c |}
    \hline
    g & \multicolumn{4}{|c|}{Data 1} & \multicolumn{4}{|c|}{Data 2} \\ \cline{2-9}
      & $\gamma_{0}$ & $\gamma_{a}$ & $\chi_{\rm{min}}^2$ & $\chi_{\rm{red},\rm{min}}^2$ & $\gamma_{0}$ & $\gamma_{a}$ & $\chi_{\rm{min}}^2$ & $\chi_{\rm{red},\rm{min}}^2$ \\ \hline
    M1 & 0.73 $\pm$ 0.05 & 0.82 $\pm$ 0.22 &27.65 & 2.77 & 0.72 $\pm$ 0.05 & 0.81 $\pm$ 0.22 &28.67 & 2.39\\ \hline
    M2 & 0.70 $\pm$ 0.05 & 0.67 $\pm$ 0.26 &40.38 & 4.04 & 0.69 $\pm$ 0.05 & 0.65 $\pm$ 0.26 &41.25 & 3.44\\ \hline
    M3 & 0.96 $\pm$ 0.09 & -0.65 $\pm$ 0.19&39.09& 3.91 & 0.95 $\pm$ 0.09 & -0.64 $\pm$ 0.19&39.88 & 3.32 \\ \hline
    M4 & 0.72 $\pm$ 0.05 & 0               &48.13 & 4.81 & 0.72 $\pm$ 0.05 &  0              &48.70 & 3.75 \\ \hline
    \end{tabular}
    \caption{$\gamma_{0}$ and $\gamma_{a}$ obtained from $\chi^2$ minimization of $g_{\obs}$ for the different models with Data 1 and 2. }
    \label{table4}
    \end{table}
\end{center}

\begin{figure}
\centering
\vspace{1.5cm}
\begin{tabular}{ccc}
\epsfig{file=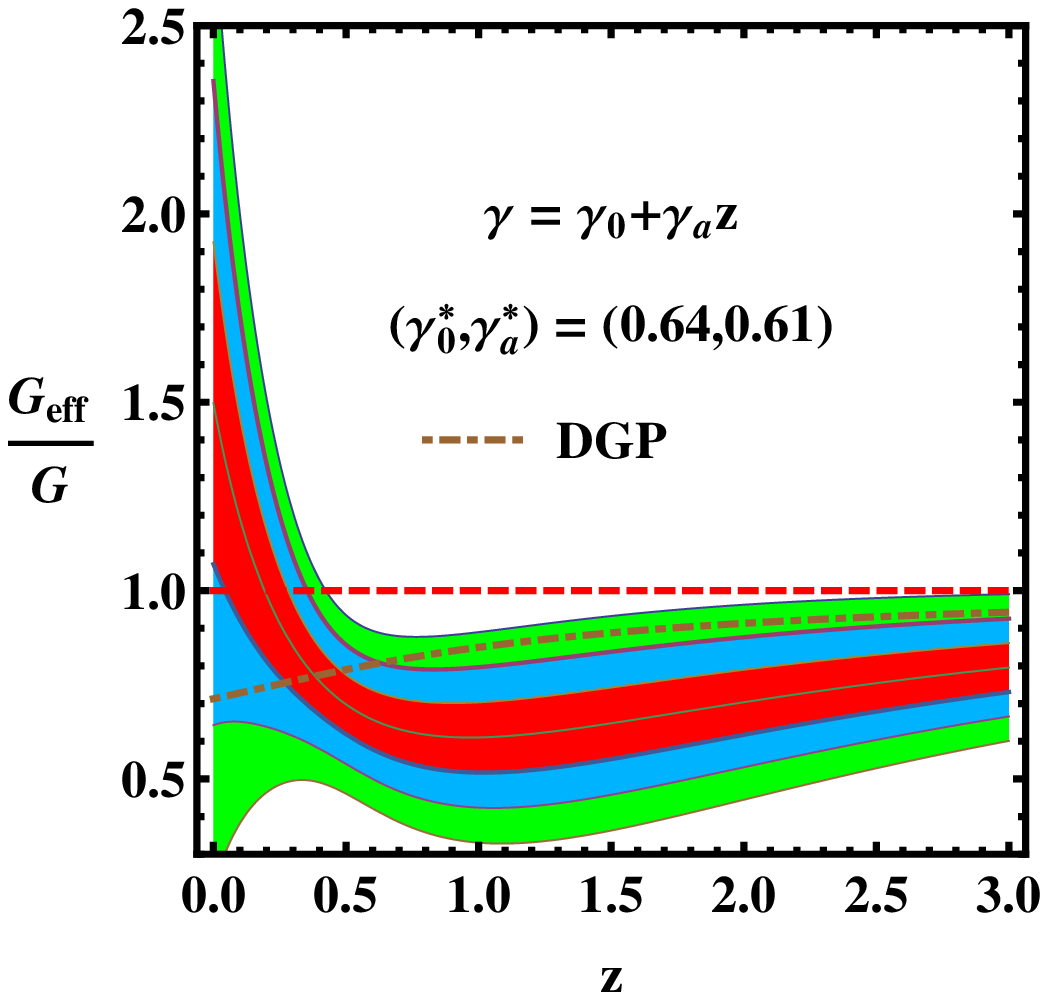,width=0.3\linewidth,clip=} &
\epsfig{file=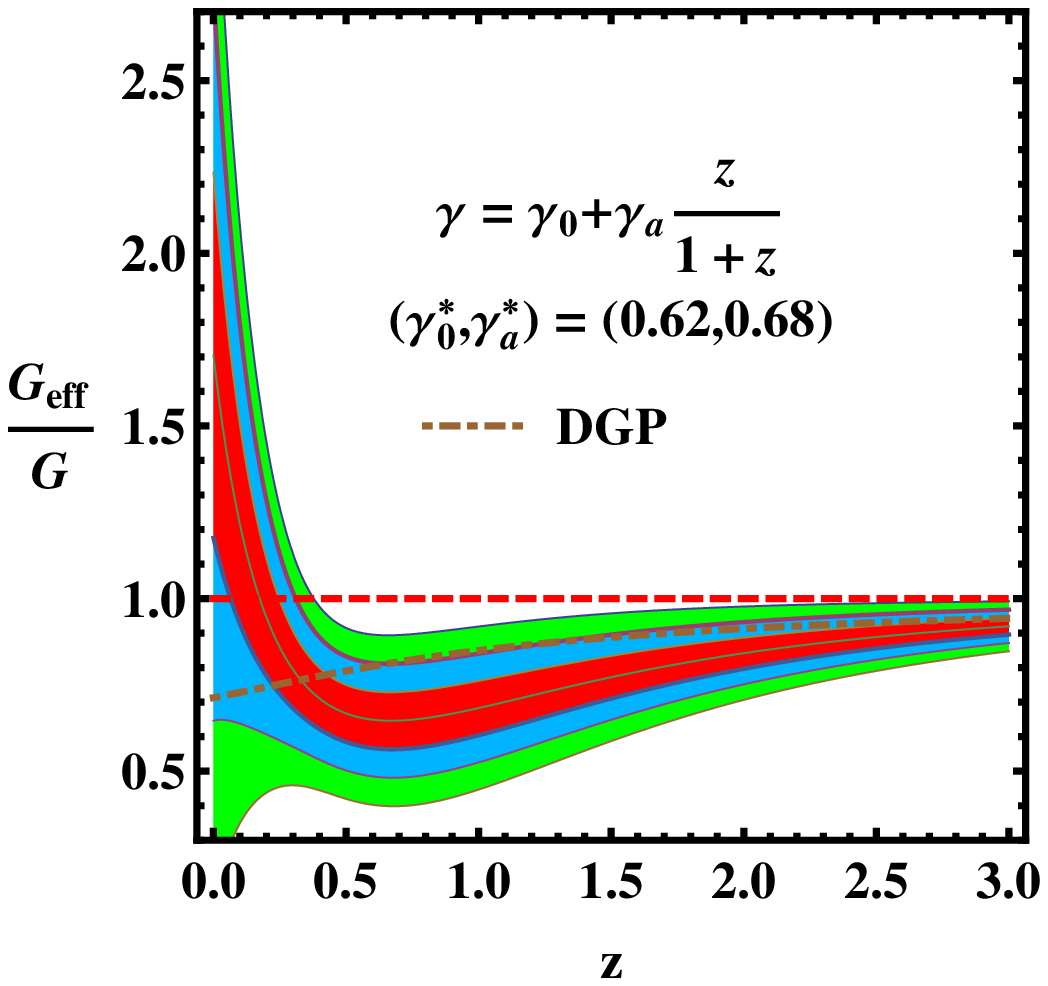,width=0.3\linewidth,clip=} &
\epsfig{file=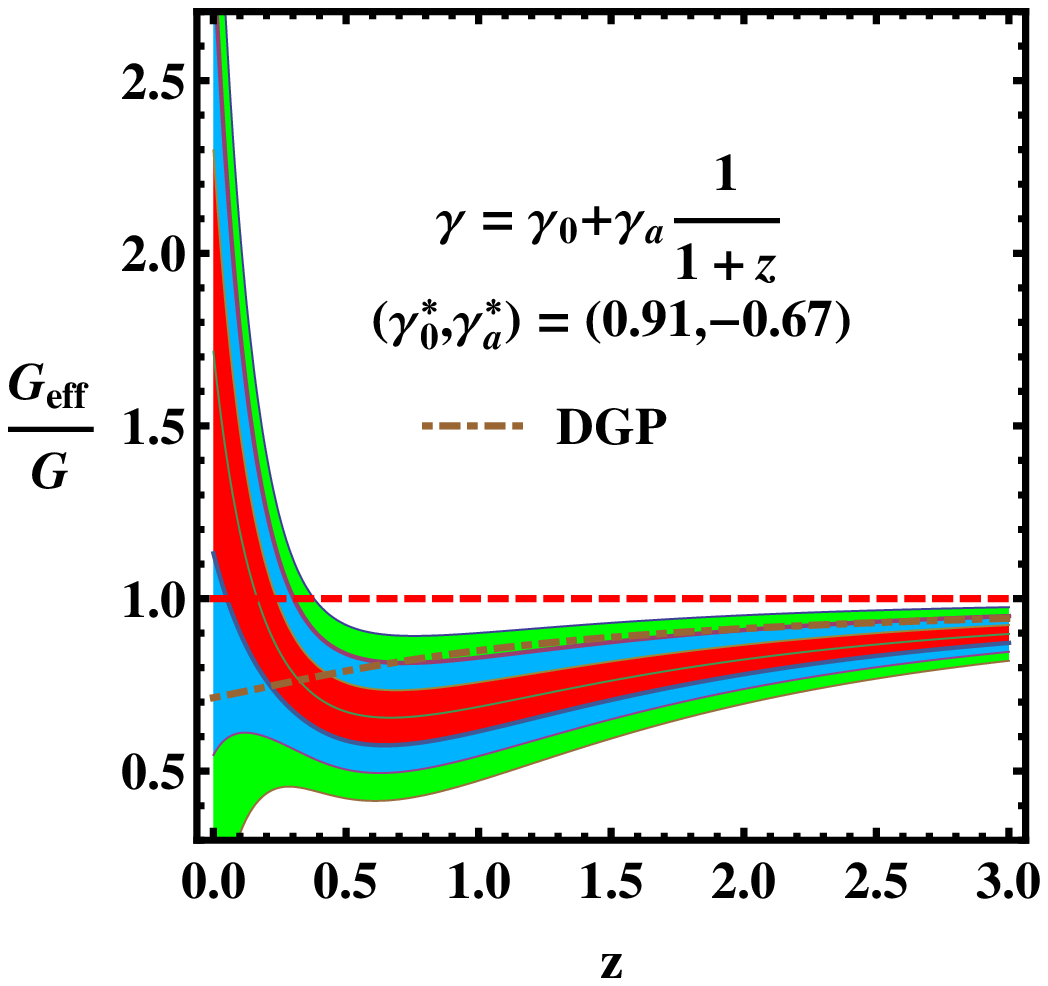,width=0.3\linewidth,clip=} \\
\epsfig{file=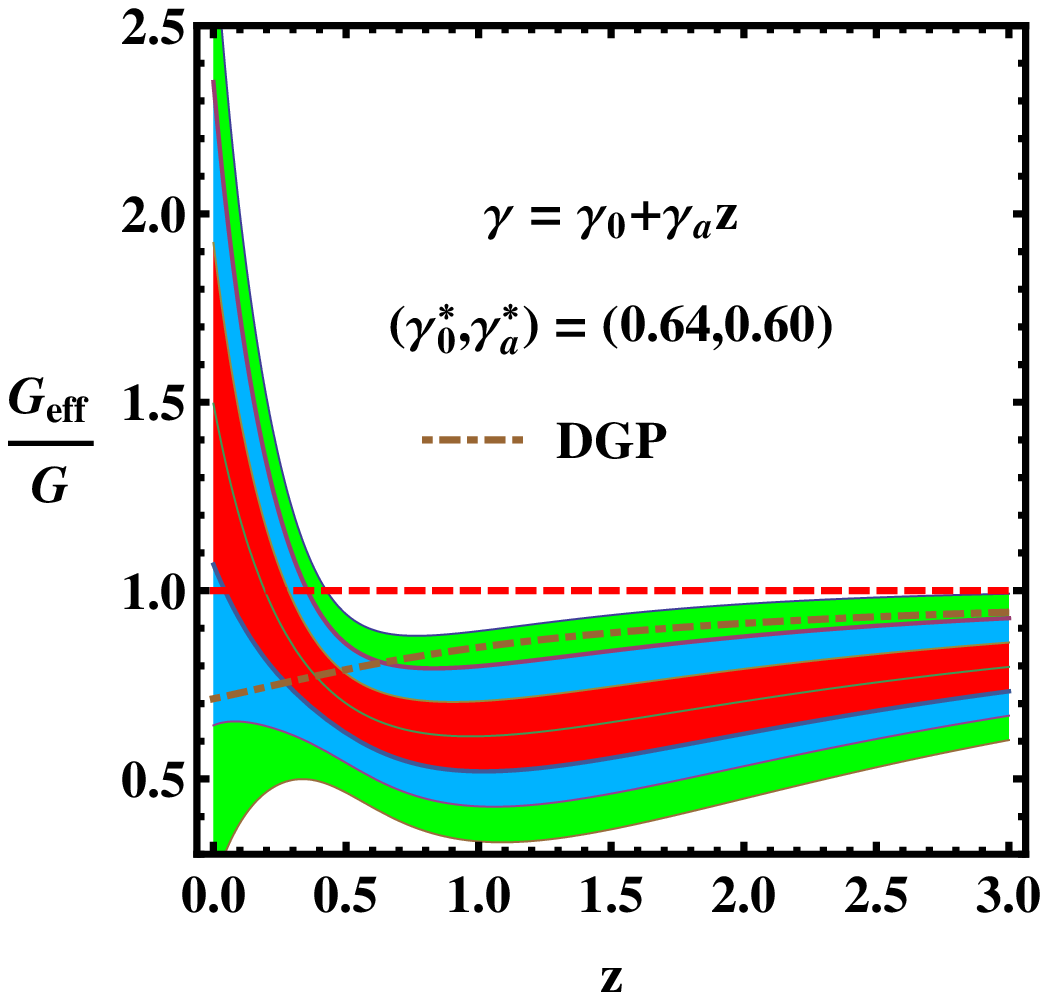,width=0.3\linewidth,clip=} &
\epsfig{file=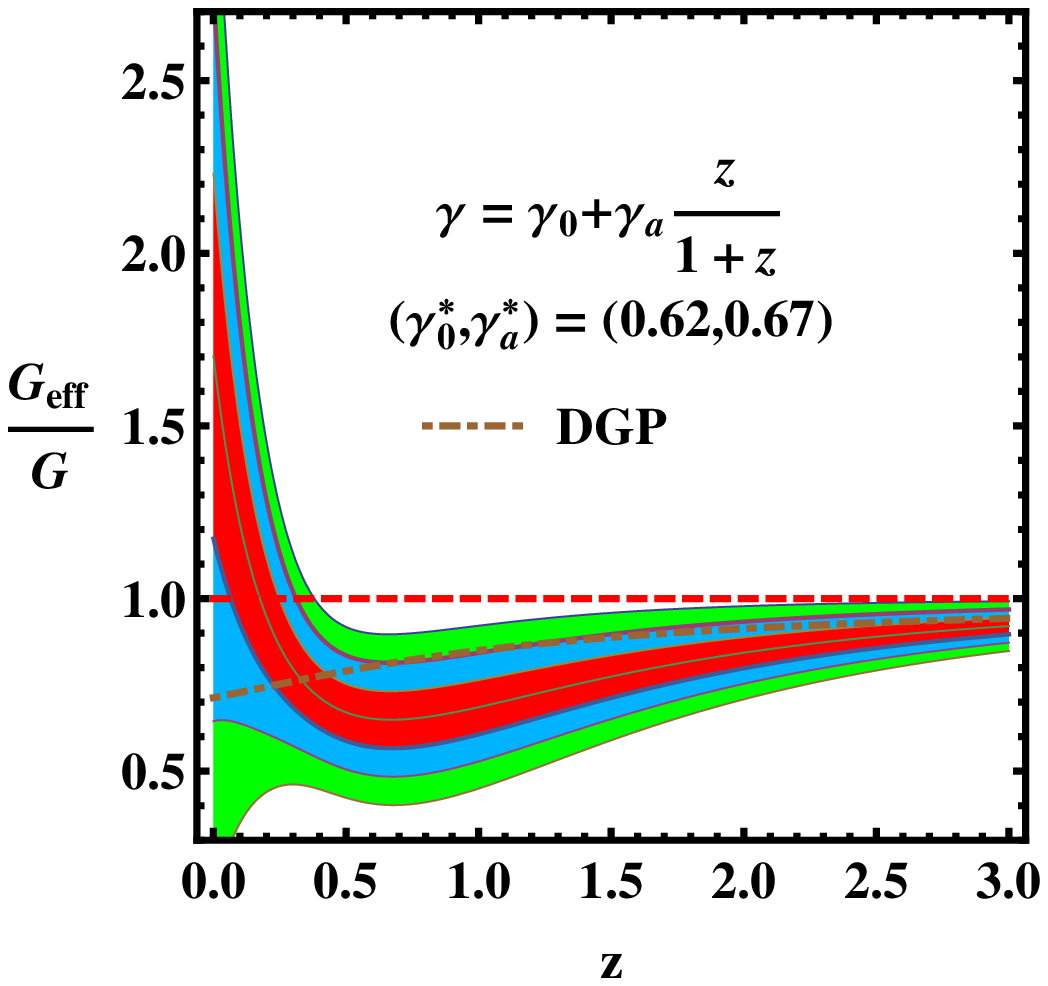,width=0.3\linewidth,clip=} &
\epsfig{file=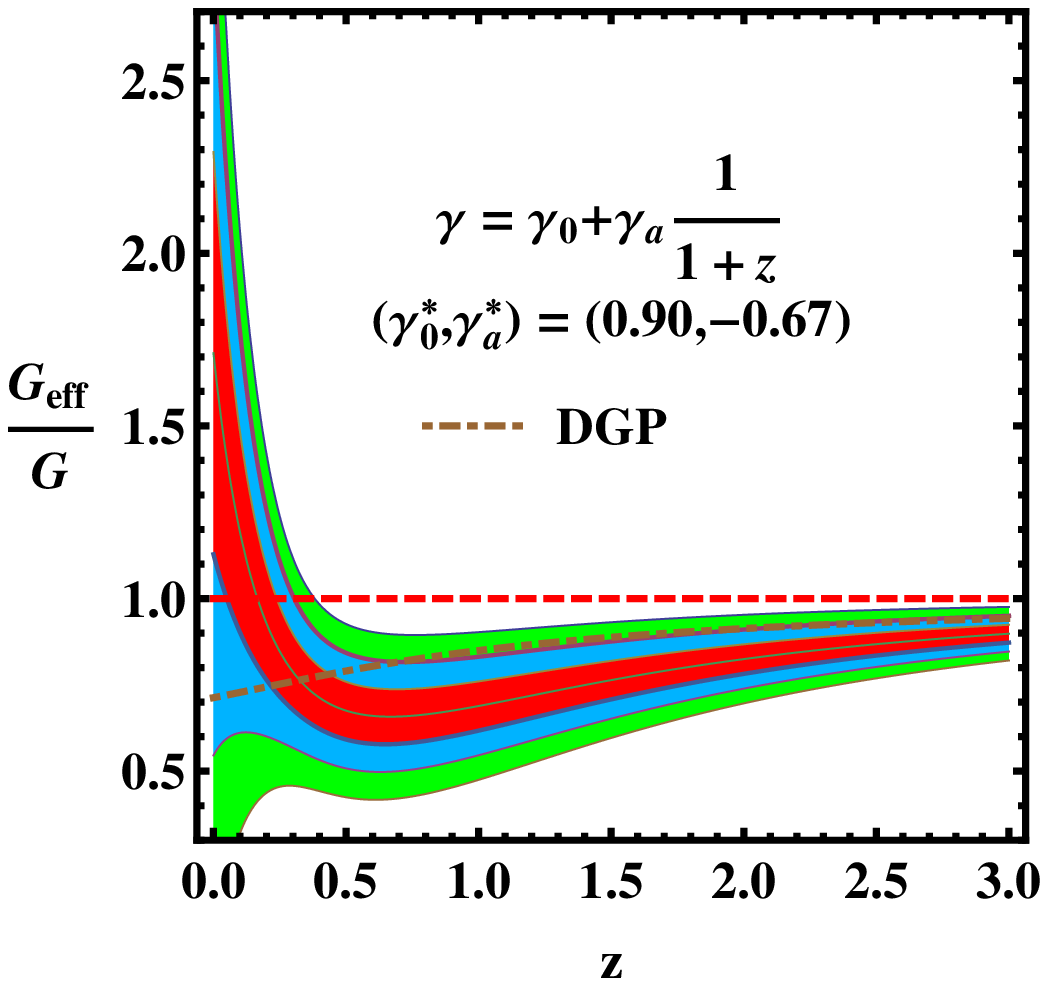,width=0.3\linewidth,clip=} \\
\end{tabular}
\vspace{-0.5cm}
\caption{ a) The best fit form of $\Geff / \GN$ for 1-$\sigma$ (darker shaded regions), 2-$\sigma$ (dark shaded regions), and 3-$\sigma$ (light shaded regions) errors in the case of Model 1 (first column), Model 2 (second column), and Model 3 (third column). The first row is obtained with using Data 3. b) Same best fit form of $\Geff / \GN$ the same models with using Data 4.} \label{fig4}
\end{figure}

Also, we repeat the $\chi^2$ minimization of $g_{\res}$ and the results are shown in Table \ref{table5} and Fig. \ref{fig4}. M1 provides $(\gamma_0, \gamma_a) = (0.64 \pm 0.05, 0.61 \pm 0.23)$ with $\chi_{\rm{min}}^2 /dof = 4.46/10$ for Data 3 and $(\gamma_0, \gamma_a) = (0.64 \pm 0.05, 0.60 \pm 0.23)$ with $\chi_{\rm{min}}^2 /dof = 5.41/12$ for Data 4, respectively. For M2, we obtain $(\gamma_0, \gamma_a) = (0.62 \pm 0.05, 0.68 \pm 0.26)$ with $\chi_{\rm{min}}^2 /dof = 5.83/10$ and $(\gamma_0, \gamma_a) = (0.62 \pm 0.05, 0.67 \pm 0.26)$ with $\chi_{\rm{min}}^2 /dof = 6.77/12$ for Data 3 and 4, respectively. For M3, $(\gamma_0, \gamma_a) = (0.91 \pm 0.08, -0.67 \pm 0.20)$ with $\chi_{\rm{min}}^2 /dof = 5.81/10$ and $(\gamma_0, \gamma_a) = (0.90 \pm 0.08, -0.67 \pm 0.20)$ with $\chi_{\rm{min}}^2 /dof = 6.74/12$ for Data 3 and 4, respectively. From M4, we obtain $\gamma_0 = 0.69 \pm 0.05$ with $\chi_{\rm{min}}^2 /dof = 13.20/10$ and $\gamma_0 = 0.69 \pm 0.05$ with $\chi_{\rm{min}}^2 /dof = 14.05/12$ for Data 3 and 4, respectively. Again, M1 is the best fit and M4 is the worst one. In Fig. \ref{fig4}, we show the 1-$\sigma$ (inner darkest shaded regions), 2-$\sigma$ (dark shaded regions) and 3-$\sigma$ (light shaded ones) best fit form contours of $g(z)$ for M1, M2, and M3 with Data 3 in the first row. We also show the same contours plots of same models with Data 4 in the second row. Again, $G_{\eff}/G_{N}$ deviates from $\Lambda$CDM expected value 1 from $z \gtrsim 0.3$ for all models at 3-$\sigma$.


\begin{center}
    \begin{table}
    \begin{tabular}{ | c | c | c | c | c | c | c | c | c |}
    \hline
    g & \multicolumn{4}{|c|}{Data 3} & \multicolumn{4}{|c|}{Data 4} \\ \cline{2-9}
      & $\gamma_{0}$ & $\gamma_{a}$ & $\chi_{\rm{min}}^2$ & $\chi_{\rm{red},\rm{min}}^2$ & $\gamma_{0}$ & $\gamma_{a}$ & $\chi_{\rm{min}}^2$ & $\chi_{\rm{red},\rm{min}}^2$ \\ \hline
    M1 & 0.64 $\pm$ 0.05 & 0.61 $\pm$ 0.23 & 4.46 & 0.45 & 0.64 $\pm$ 0.05 & 0.60 $\pm$ 0.23 & 5.41 & 0.45\\ \hline
    M2 & 0.62 $\pm$ 0.05 & 0.68 $\pm$ 0.26 & 5.83 & 0.58 & 0.62 $\pm$ 0.05 & 0.67 $\pm$ 0.26 & 6.77 & 0.56\\ \hline
    M3 & 0.91 $\pm$ 0.08 & -0.67 $\pm$ 0.20 & 5.81& 0.58 & 0.90 $\pm$ 0.08 & -0.67 $\pm$ 0.20 & 6.74& 0.56 \\ \hline
    M4 & 0.69 $\pm$ 0.05 & 0               & 13.20& 1.32 & 0.69 $\pm$ 0.05 &  0              & 14.05& 1.17 \\ \hline
    \end{tabular}
    \caption{$\gamma_{0}$ and $\gamma_{a}$ obtained from $\chi^2$ minimization of $g_{\res}$ for the different models with Data 3 and 4. }
    \label{table5}
    \end{table}
\end{center}

\section{Conclusions}
\setcounter{equation}{0}

After we have compiled a currently available dataset of the growth function from the fiducial concordance $\Lambda$CDM model for the background evolution, we also re-scale the values of the observed growth function. We constrain the several phenomenologically parameterized growth index parameter $\gamma$ allowing time variation of it. Theoretically, almost the constant $\gamma$ is expected from the $\Lambda$CDM model, but the best model from  original data produces $\gamma(z) = \gamma_0 + \gamma_a z$ with the best fit values $(\gamma_0, \gamma_a) = (0.48 \pm 0.07, 0.32 \pm 0.20)$ and $(0.47 \pm 0.07, 0.34\pm 0.20)$ with and without 2 SLAQ and Lyman-$\alpha$ data, respectively. The reduced $\chi^2$ values are both 0.74. When we use the rescaled values of $f$, then the best fit values are $(\gamma_0, \gamma_a) = (0.50 \pm 0.07, 0.30 \pm 0.20)$ and $(0.50 \pm 0.07, 0.34\pm 0.20)$ with and without 2 SLAQ and Lyman-$\alpha$ data, respectively. The reduced $\chi^2$ values are 0.48 and 0.28, respectively. Even though the constant $\gamma$ model includes the $\Lambda$CDM expected value of $\gamma_{\Lambda}(\Omo = 0.273) = 0.555$ at 1-$\sigma$ level, this gives the worse fit compared to other models for all data.

We also investigate the time variation of the effective gravitational constant proposed by modified gravity theories. It is interesting to find that the effective gravitational constant is smaller than the Newton's gravitational constant $G_{N}$ for $z \gtrsim 0.2 \sim 0.3$ for all considered models at 3-$\sigma$ level. Thus, we can conclude that the current large scale structure formation shows the deviation from Einstein's General Relativity $\Lambda$CDM with 99 \% confidence level.

The current growth rate observations show the deviation from Einstein's general relativity. By using the previous data with $20 - 40$ \% accuracies and WiggleZ galaxy survey within $9 - 17$ \%, the errors in $\gamma$ parameters are about 15 \% and 60 \% for $\gamma_0$ and $\gamma_a$, respectively.

\section*{Acknowledgments}
This work was supported in part by the National Science Council, Taiwan, ROC under the
Grant NSC 98-2112-M-001-009-MY3.


\begin{thebibliography}{99}

\bibitem{09051522} S.~Lee and K.-W.~Ng, [astro-ph/0905.1522].

\bibitem{09061643} S.~Lee and K.-W.~Ng, Phys.\ Lett.\ B {\bf 688}, 1 (2010) [arXiv:0906.1643]. 

\bibitem{09072108} S.~Lee and K.-W.~Ng, Phys.\ Rev.\ D {\bf 82}, 043004 (2010) [arXiv:0907.2108]. 

\bibitem{12021637} S.~Basilakos, [arXiv:1202.1637]. 

\bibitem{12044725} F.~Beutler {\it et al.}, [arXiv:1204.4725]. 

\bibitem{08021944} L.~Guzzo {\it et al.}, Nature {\bf 451}, 541 (2008) [arXiv:0802.1944]. 

\bibitem{08070810} Y.-S.~Song and W.~J.~Percival, \JCAP {\bf 0910}, 004 (2009) [arXiv:0807.0810]. 

\bibitem{0608632} M.~Tegmark {\it et al.}, Phys.\ Rev.\ D {\bf 74}, 123507 (2006) [arXiv:astro-ph/0608632]. 

\bibitem{0612400} N.~P.~Ross {\it et al.}, Mon.\ Not.\ Roy.\ Astron.\ Soc.\ {\bf 381}, 573 (2007) [arXiv:astro-ph/0612400]. 

\bibitem{0612401} J.~da Angela {\it et al.}, Mon.\ Not.\ Roy.\ Astron.\ Soc.\ {bf 383}, 565 (2008)
[arXiv:astro-ph/0612401].

\bibitem{0404600} M.~Viel, M.~G.~Haehnelt, V.~Springel, Mon.\ Not.\ Roy.\ Astron.\ Soc.\ {\bf 354}, 684 (2004) [arXiv:astro-ph/0404600].

\bibitem{0407377} P.~McDonald {\it et al.}, Astrophys.\ J.\ {\bf 635}, 761 (2005) [arXiv:astro-ph/0407377]. 

\bibitem{11042948} C.~Blake {\it et al.}, Mon.\ Not.\ Roy.\ Astron.\ Soc.\ {\bf 415}, 2876 (2011) [arXiv:1104.2948]. 

\bibitem{12043674} C.~Blake {\it et al.}, [arXiv:1204.3674]. 

\bibitem{07101510} D.~Polarski and R.~Gannouji, Phys.\ Lett.\ B {\bf 660}, 439 (2008) [arXiv:0710.1510]. 

\bibitem{10043086} J.~Dossett {\it et al.}, \JCAP {\bf 1004}, 022 (2010) [arXiv:1004.3086]. 

\bibitem{12036724} S.~Basilakos and A.~Pouri, [arXiv:1203.6724]. 

\end{thebibliography}
\end{document}